\documentclass[twocolumn,twocolappendix]{aastex631}

\usepackage{graphicx}
\usepackage{multirow}
\usepackage[sort&compress]{natbib}
\usepackage{hyperref}
\usepackage{rotating}
\usepackage{graphicx}
\usepackage{supertabular}
\usepackage{longtable}
\usepackage{CJK}
\usepackage[T1]{fontenc}
\usepackage[utf8]{inputenc}
\usepackage{tikz}
\usepackage{threeparttable}
\usepackage{xcolor}

\newcommand\feh{{\rm [Fe/H]}}
\newcommand\logg{{\rm log}\,g}
\newcommand\teff{T_{\rm eff}}
\newcommand\tefflm{\rm T_{eff, LAMOST}}

\newcommand\fehlm{{\rm [Fe/H]_{LAMOST}}}
\newcommand\logglm{{\rm log}\,g_{\rm LAMOST}}

\newcommand\ebv{E(B$-$V)}
\newcommand\sfdebv{\rm E(B-V)_{SFD}}
\newcommand\gaia{Gaia}
\usepackage{amsmath}
\usepackage{soul}

\begin{document}
\begin{CJK*}{UTF8}{gbsn}
\title{An all-sky 3D dust map Based on Gaia and LAMOST}
\correspondingauthor{Haibo Yuan (苑海波)}
\email{yuanhb@bnu.edu.cn}

\author[0000-0002-4878-1227]{Tao Wang (王涛)}
\affiliation{Institute for Frontiers in Astronomy and Astrophysics,
Beijing Normal University, Beijing, 102206, China; }
\affiliation{School of Physics and Astronomy, Beijing Normal University No.19, Xinjiekouwai St, Haidian District, Beijing, 100875, China; }

\author[0000-0003-2471-2363]{Haibo Yuan (苑海波)}
\affiliation{Institute for Frontiers in Astronomy and Astrophysics,
Beijing Normal University, Beijing, 102206, China; }
\affiliation{School of Physics and Astronomy, Beijing Normal University No.19, Xinjiekouwai St, Haidian District, Beijing, 100875, China; }

\author[0000-0003-2472-4903]{Bingqiu Chen (陈丙秋)}
\affil{South-Western Institute for Astronomy Research, Yunnan University, Kunming 650500, China;} 

\author[0000-0002-5818-8769]{Maosheng Xiang (向茂盛)}
\affiliation{National Astronomical Observatories, Chinese Academy of Sciences, 20A Datun Road, Chaoyang District, Beijing, China}
\affiliation{Institute for Frontiers in Astronomy and Astrophysics,
Beijing Normal University, Beijing, 102206, China; }

\author[0000-0003-1863-1268]{Ruoyi Zhang (张若羿)}
\affiliation{Institute for Frontiers in Astronomy and Astrophysics,
Beijing Normal University, Beijing, 102206, China; }
\affiliation{School of Physics and Astronomy, Beijing Normal University No.19, Xinjiekouwai St, Haidian District, Beijing, 100875, China; }
\affil{Max Planck Institute for Astronomy, Königstuhl 17, Heidelberg D-69117, Germany}

\author[0000-0002-1259-0517]{Bowen Huang (黄博闻)} 
\affiliation{Institute for Frontiers in Astronomy and Astrophysics,
Beijing Normal University, Beijing, 102206, China; }
\affiliation{School of Physics and Astronomy, Beijing Normal University No.19, Xinjiekouwai St, Haidian District, Beijing, 100875, China; }

\author[0009-0007-5610-6495]{Hongrui Gu (顾弘睿)}
\affiliation{CAS Key Laboratory of Optical Astronomy, National Astronomical Observatories, Chinese Academy of Sciences, Beijing 100101, People's Republic of China;}
\affiliation{School of Astronomy and Space Science, University of Chinese Academy of Sciences, Beijing 100049, People's Republic of China;}

\author[0009-0000-0467-3171]{Shuaicong Wang (王帅聪)}
\affiliation{National Astronomical Data Center, Beijing 100101, People's Republic of China}
\affiliation{National Astronomical Observatories, Chinese Academy of Sciences, 20A Datun Road, Chaoyang District, Beijing, China}

\author[0000-0003-4108-4979]{Jiawei Li (李佳伟)}
\affiliation{State Key Laboratory of Quantum Optics Technologies and Devices, 
Shanxi University, Taiyuan, Shanxi, 030006, China; }

\begin{abstract}

We present a comprehensive 3D dust reddening map covering the entire Milky Way, constructed by combining reddening estimates based on LAMOST low-resolution spectra (E(B$-$V)$_{\rm LAMOST}$) with those derived from $\gaia$ XP spectra (E(B$-$V)$_{\rm XP}$), along with revised $\gaia$ distances. E(B$-$V)$_{\rm LAMOST}$ values of $\sim$ 4.6 million unique sources were obtained with the standard-pair analysis using LAMOST DR11 stellar parameters and synthesized $B/V$-band photometry from $\gaia$ XP spectra, showing a typical precision of $\sim$ 0.01 mag. The E(B$-$V)$_{\rm XP}$ from the catalog of \citet{zhang2023}, which was derived using forward modeling of $\gaia$ XP spectra, were cross-validated with E(B$-$V)$_{\rm LAMOST}$, leading to the selection of $\sim$ 150 million high-reliability measurements. The combined dataset achieves a median precision of $\sim$ 0.03 mag for E(B$-$V). To model the reddening--distance relationship along various lines-of-sight, we implemented a parametric approach that accounts for contributions from the local bubble, diffuse interstellar-medium, and multiple potential molecular clouds. The sky was adaptively partitioned based on stellar density, resulting in angular resolutions ranging from 3.4$^{\prime}$ to 58$^{\prime}$, with about half of the sky having a resolution better than 6.9$^{\prime}$. The reddening precision of our 3D map for individual stars reaches $\sim$ 0.01 mag in most regions at $|b| > 20^\circ$, but degrades to 0.01 -- 0.05 mag at $|b| < 20^\circ$. The map reaches a maximum distance of 3 -- 5 kpc in high-extinction regions with $|b| < 5^\circ$, and extends to 10 -- 15 kpc elsewhere. An interactive platform and Python package have been developed for utilization of the 3D dust map.

\end{abstract}


\section{Introduction} \label{sec:intro}

Interstellar dust is widespread throughout the Milky Way (MW), constituting only 1\% of the mass of the interstellar medium but absorbing approximately 30\% of the starlight \citep{2003ARA&A..41..241D}, significantly impacting the observed properties of celestial objects. Interstellar dust absorbs and scatters ultraviolet (UV), optical, and near-infrared (NIR), re-emitting in the mid-infrared (MIR) to far-infrared (FIR), leading to a reduction in luminosity and a shift in color, a phenomenon known as extinction and reddening. In UV, optical, and NIR astronomy, dust acts as an extinction foreground, requiring correction for the effects of extinction and reddening, particularly in the context of objects within the MW. In cosmology and FIR astronomy, dust acts as an emitting foreground, necessitating the removal of dust emission. In the study of Galactic structure and star formation, dust itself is a key research subject. Therefore, a reliable extinction map provides a valuable tool for depicting the spatial distribution of interstellar dust and its attenuation of starlight, calibrating emission-based dust maps, studying Galactic structure, and determining the distances to objects responsible for the reddening effect.

Dust can be mapped both through extinction and emission. Traditionally, the commonly used 2D extinction maps are typically constructed by modeling the dust emission spectral energy distribution to obtain a dust column density map, which is then calibrated using the extinction of celestial objects to convert the dust column density into extinction. Examples include the extinction map of \cite{SFD1998} (hereafter SFD) based on FIR data and the \cite{Planck2014} map based on FIR and microwave data. These maps provide integrated extinction along different lines-of-sight, and have been widely used in extragalactic astronomy and studies of field stars beyond the MW's dust disk. However, this emission-based method can only recover the 2D distribution of dust on the sky but not its distance distribution, which poses a significant challenge for studying objects within the MW, particularly those located within the dust disk. Moreover, 2D extinction maps only provide projections of dust in the sky, failing to reveal the true three-dimensional distribution of interstellar dust. Therefore, to achieve more accurate extinction corrections and better characterize the structure of the MW, there is a need of transition from two-dimensional extinction maps to three-dimensional extinction maps.

Constructing a 3D extinction map of the MW is typically based on estimates of extinction or reddening from sources distributed across the sky, which requires extinction and distance information for celestial objects. However, there is a degeneracy between extinction and distance, which complicates the interpretation. Early studies often relied on model assumptions to break this degeneracy (e.g., \citealt{2001ApJ...556..181D,2003A&A...409..205D,2006A&A...453..635M,2013A&A...550A..42C}), but such approaches are typically associated with significant systematic errors, limiting the precision and reliability of the results. Recent large-scale multi-band photometric and spectroscopic surveys, combined with the extensive astrometric data from the $\gaia$ mission, have significantly enhanced our ability to obtain accurate extinction and distance measurements.  Spectroscopic surveys provide precise extinction data, while $\gaia$'s astrometric measurements deliver a vast number of accurate stellar distances.  These advancements have paved the way for constructing comprehensive three-dimensional extinction maps of the entire sky.

In this context, numerous recent studies have developed various methods for constructing three-dimensional extinction maps. For example, \cite{Green2019} combined multi-band photometric data from Pan-STARRS1 and 2MASS with astrometric measurements from $\gaia$ DR2 to calculate the distances and reddening values of nearly 800 million stars, resulting in a three-dimensional extinction map that covers the northern sky (declination > −30$^\circ$) and probes distances up to approximately 6 kpc; \cite{2019MNRAS.483.4277C} integrated multi-band photometric data from $\gaia$, 2MASS, and WISE with $\gaia$ DR2 astrometric data and utilized spectroscopic data from LAMOST, APOGEE, and SEGUE to train machine learning algorithms, enabling them to estimate the color excess values for 56 million stars and construct a three-dimensional extinction map of the Galactic disk extending to distances of 6 kpc; \cite{2021ApJ...906...47G} employed multi-band photometric data from SMSS, 2MASS, WISE, and $\gaia$, along with $\gaia$ DR2 parallax measurements, to estimate r-band extinction values for approximately 19 million stars, creating a three-dimensional extinction map of the southern sky; Recently, \cite{2024ApJS..272...20A} utilized the extensive low-resolution XP spectra from $\gaia$ DR3 to construct a three-dimensional extinction map that covers the entire sky and probes distances up to approximately 3 kpc. In addition to these studies, other researchers have proposed various methods for constructing three-dimensional extinction maps (e.g., \citealt{2022AA...664A.174V,2022AA...661A.147L,2024MNRAS.532.3480D} etc.).

LAMOST parameters allow for precise measurements of intrinsic stellar colors and, when combined with $\gaia$'s Blue Photometer and Red Photometer (BP/RP, hereafter XP) synthesized magnitudes, enable accurate determination of stellar extinction. However, LAMOST is restricted to the northern hemisphere and encompasses approximately five million unique stars. With the release of $\gaia$ DR3, we now have access to 220 million XP spectra covering the entire sky. 

\cite{zhang2023} apply a forward model to estimate stellar atmospheric parameters, distances, and extinctions for 220 million stars, utilizing XP spectra from Gaia DR3.
Therefore, our study leverages LAMOST and $\gaia$ DR3 XP spectral data, along with $\gaia$ astrometry, to map the dust distribution across the entire sky, including the $|b| < 5^\circ$ region, which reaches up to 3 -- 5 kpc, and the $|b| > 5^\circ$ region, which extends to 10 -- 15 kpc. The vast number of precise extinction measurements allows us to enhance the angular resolution of our maps, achieving resolutions ranging from 3.4 to 58 arcminutes while maintaining high precision. Although previous studies have produced extinction maps using photometric data with similar spatial coverage, the extinction precision of photometric data is generally lower than that of spectroscopic data. For example, \cite{Green2019} reported a typical $E(B-V)$ uncertainty of about 0.07 mag for individual stars using SED fitting with multi-band photometry. In comparison, the Gaia XP-based $E(B-V)$ measurements used in this work have typical uncertainties of 0.03 mag, while the typical uncertainties for LAMOST spectroscopic measurements are about 0.01 mag. Therefore, extinction maps constructed from photometric data generally have lower precision than those based on spectroscopic data. In addition, by employing a physically motivated parametric model for the line-of-sight direction, we have further enhanced the interpretability of our results. By combining all these features and advantages, our map provides a valuable opportunity for precise reddening correction and in-depth studies of the dust and molecular cloud structures in the MW.

The structure of this paper is organized as follows. In Section \ref{sec:data}, we derive the \ebv for approximately 4.6 million unique stars using the stellar atmospheric parameters from LAMOST DR11, achieving a typical precision of $\sim$ 0.01 mag. 
Furthermore, we use the LAMOST \ebv~ as a reference and selected sample stars of \cite{zhang2023}, filtering a total of $\sim$ 150 million sources, which are then merged into the final dataset.
In Section \ref{sec-Method}, we describe the process of dividing the sky into different lines-of-sight and the methodology we employed to model each line-of-sight and construct the 3D extinction map. Section \ref{sec-Result} presents our results and compares the 3D extinction map with previously maps. In Section \ref{sec-specification}, we provide details on how to access our data products. Finally, Section \ref{sec-Summary} offers a summary of our work.

\section{Data} \label{sec:data}
\subsection{LAMOST Data}\label{sec:LAMOST- Data}

The Large Sky Area Multi-Object Fiber Spectroscopy Telescope (LAMOST; \citealt{2012RAA....12.1197C,2012RAA....12..723Z,2014IAUS..298..310L}), a 4 m quasi-meridian reflective Schmidt telescope, is located at the Xinglong Observatory of the National Astronomical Observatories of China.  On March 28, 2024, LAMOST released the LAMOST Data Release 11 (version 1.0), hereafter referred to as LAMOST DR11 v1.0 (\url{https://www.lamost.org/dr11/v1.0}), which includes over 25.12 million spectra. 

This work utilizes the stellar parameters from the LAMOST DR11 v1.0's Low-Resolution Spectroscopic Survey (LRS), specifically the LAMOST LRS Stellar Parameter Catalogue of A, F, G, and K Stars, for the standard-pair algorithm. This catalog contains stellar atmospheric parameters for 7,774,147 spectra, such as effective temperature ($\teff$), surface gravity ($\logg$), and metallicity ($\feh$).
These parameters have been derived by the LAMOST Stellar Parameter Pipeline with typical errors of approximately 150 K, 0.25 dex, and 0.15 dex, respectively \citep{2011RAA....11..924W,2015RAA....15.1095L}. It is important to note that our methodology focuses on the relative ranking of stars within the Teff, log g, and [Fe/H] parameter space, rendering systematic uncertainties less critical.  Furthermore, the uncertainties in these parameters are primarily systematic, while the internal errors are significantly smaller (e.g., \citealt{2021ApJ...909...48N}).  Note also that the systematic errors are unaffected by stellar extinction, as the LAMOST stellar parameters are determined based on normalized spectra.

In the LAMOST DR11, approximately 20\% of the sources were observed multiple times. To ensure the precision of our subsequent data fitting,we retained only the measurement with the highest signal-to-noise ratio (SNR) for each duplicated source,
resulting in approximately 5.5 million unique sources. Specifically, for the duplicate sources in LAMOST DR11, we retained the dataset with the highest SNR. We matched the LAMOST DR11 sources, after removing duplicates, to the $\gaia$ DR3 Synthetic Photometry catalog \citep{2023A&A...674A..33G} using the $\gaia$ source identifier (source\_id) provided in LAMOST DR11 to obtain synthetic magnitudes in the B and V bands. Due to the incomplete coverage of LAMOST sources by $\gaia$'s XP spectra, we matched $\sim$ 4.8 million sources 
which were then used in the standard-pair algorithm.

\subsection{Reddening Measurements of LAMOST Stars with the Standard-Pair Algorithm}\label{sec:Standard-Pair}

To obtain accurate reddening values \ebv, the intrinsic color of the star is first obtained through the standard-pair algorithm \citep{2013MNRAS.430.2188Y}, which is based on the  assumption that "stars with the same atmospheric parameters have the same intrinsic color."  

We select stars with very low extinction ($\sfdebv$ < 0.01 mag) from the SFD map as a control sample. We assume that the colors of these control stars, after correcting for $\sfdebv$, represent the intrinsic colors of stars with identical atmospheric parameters, i.e., (B$-$V)$_{\rm intrinsic}$ = (B$-$V)$_{\rm observed}$ $-$ $\sfdebv \times 0.86$  (Here we apply a scale factor of 0.86 to SFD to correct for the systematic difference from stellar locus measurements \citep{2010ApJ...725.1175S,2011ApJ...737..103S,2013MNRAS.430.2188Y}.
Given the large number of stars with precise stellar parameters in the LAMOST catalog, for most target stars, it is generally possible to identify a group of nearly unreddened stars with similar stellar parameters ($\tefflm$, $\logglm$, and $\fehlm$). In regions with dense parameter coverage, sufficient comparison stars can be typically found within a small range of parameters. However, in regions with sparse parameter coverage, a broader tolerance in the parameter space must be considered to ensure an adequate sample of comparison stars.

The criteria for selecting the control stars follow those outlined in Section 3 of \cite{2023ApJS..264...14Z}. To determine the intrinsic color of the target star, we perform a simple linear fit for each target star using the following formula:$ \mathrm{(B-V)}_0 = a \times T_{\mathrm{eff, LAMOST}} + b \times \log g_{\mathrm{LAMOST}} + c \times [\mathrm{Fe/H}]_{\mathrm{LAMOST}} + d,$ where $a$, $b$, $c$, and $d$ are free parameters. Finally, after obtaining the unique best fit parameters $a$, $b$, $c$, and $d$ for each star, we substitute the stellar parameters of the target star into this linear equation. The resulting (B$-$V)$_{0}$ is the intrinsic color of the target star. To ensure the quality of the fit, we limit the number of control stars to no less than 10.

\begin{figure}
    \centering
    \includegraphics[width=1\linewidth]{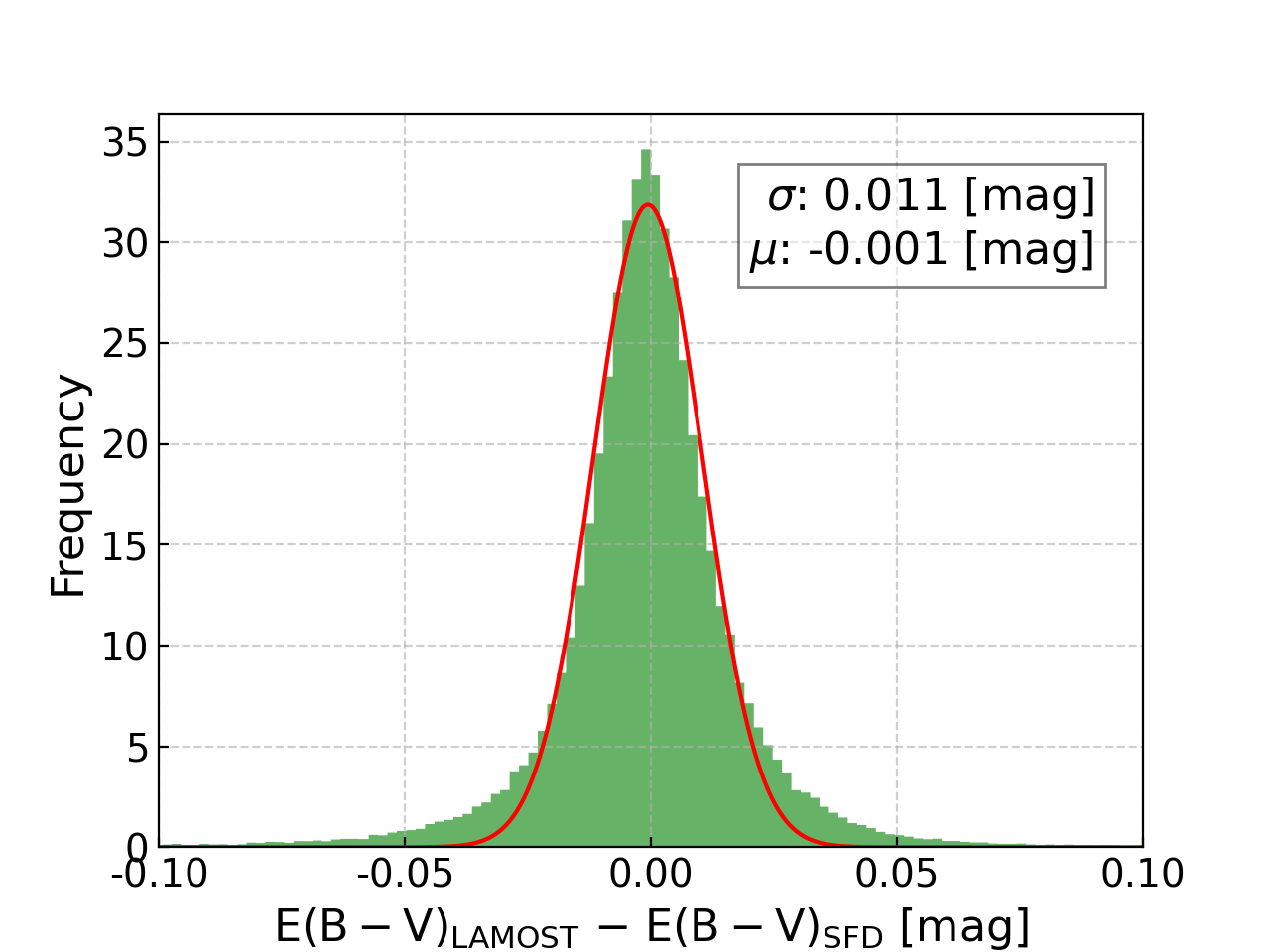}
    \caption{Mean and standard deviation of the differences between the two independent measurements of extinction (\ebv$_{\rm{LAMOST}}$ $-$ $\sfdebv$). }
    \label{fig:Difference_between_sample_and_SFD}
\end{figure}

To further refine the control sample, we first perform a "self-pairing" process on all stars in the control sample by treating each star as a target sample. For each star in the control sample, we obtain its intrinsic color using two independent methods: one is based on the stellar parameters and estimated using the standard-pair algorithm, denoted as \ebv$_{\rm LAMOST}$; the other is obtained from the SFD map, denoted as $\sfdebv$. By calculating the difference between these two measurements (\ebv$_{\rm LAMOST}$ $-$ $\sfdebv$) and fitting it with a Gaussian distribution, we find a standard deviation $\sigma \approx 0.011$ mag. This standard deviation reflects the extinction precision achieved using the standard-pair algorithm, as shown in Figure \ref{fig:Difference_between_sample_and_SFD}. We iteratively exclude samples that deviate by more than 3-$\sigma$ range from the control sample, resulting in a final control sample of $\sim$ 0.11 million stars, representing about 2.6\% of the overall target sample. 

\begin{figure*}
    \centering
    \includegraphics[width=1\linewidth]{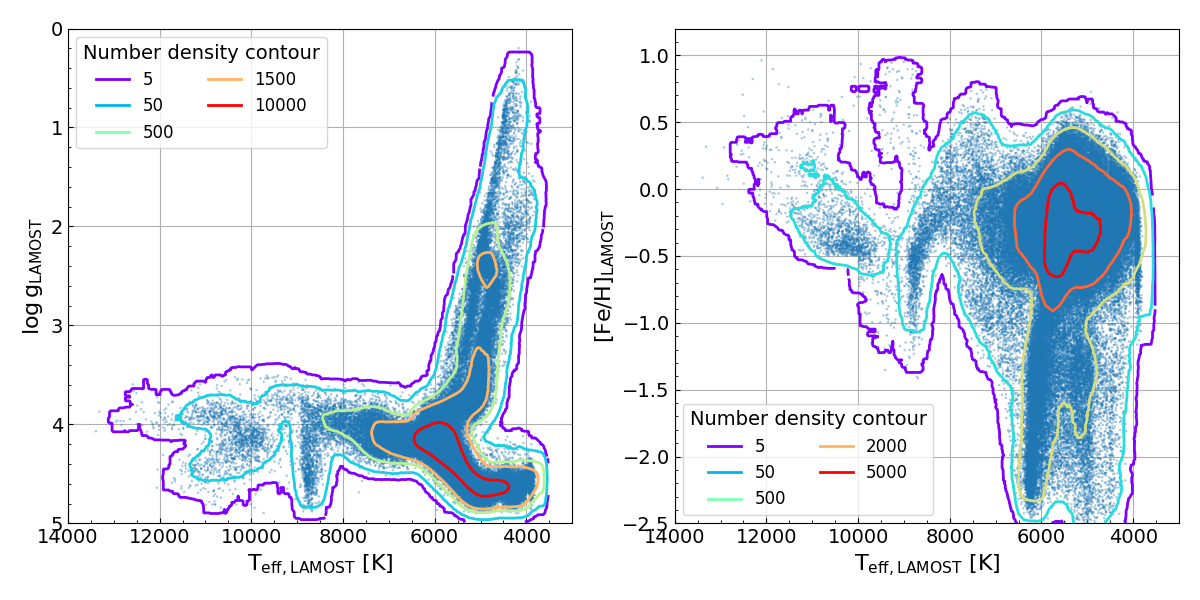}
    \caption{$\tefflm$ vs. $\logglm$ (left panel) and $\tefflm$ vs. $\fehlm$ (right panel) distributions of the target and control samples. The blue dots represent the control sample stars, and the density contours represent the target sample.}
    \label{fig:Distribution_parameter_space}
\end{figure*}

Figure \ref{fig:Distribution_parameter_space} shows the distribution of the target sample and the control sample in the stellar parameter space, demonstrating that the control sample covers the parameter space of the target sample. Due to the relaxed selection criteria for metal-poor stars in the control sample, the giant branch of the control stars exhibits a dual structure (the increase in metal-poor stars leads to the presence of an upper branch), corresponding to two groups of stars with different metallicities. The dense region between the two branches consists of red-clump stars.

\begin{figure*}
    \centering
    \includegraphics[width= 1\linewidth]{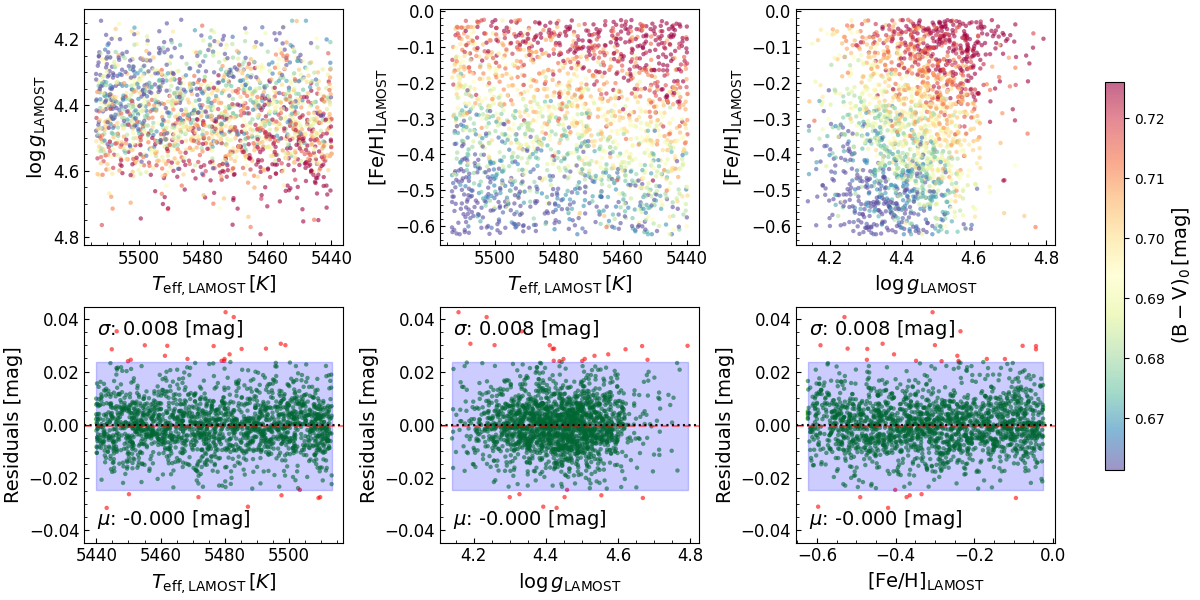}
    \caption{An example of the results for a target star using the standard-pair algorithm. The top three panels show the distribution of control stars in the parameter space, with the colors encoding the intrinsic colors of the control stars. The bottom three panels display the residuals from the linear fit, with the blue shading and red dashed lines indicating the 3-$\sigma$ range and the mean ($\mu$) of the residuals, respectively. The gray dashed line marks the zero level.}
    \label{fig:Targetl_sample_fit_example}
\end{figure*}

We used the entire sample of $\sim$ 5 million stars as the target sample and applied the aforementioned standard-pair algorithm by selecting stars from the control sample. For 99.6\% of the target stars, sufficient control stars were found to perform the linear fit. An example of this fitting is shown in Figure \ref{fig:Targetl_sample_fit_example}. The top three subplots illustrate the distribution of intrinsic color in the parameter space constituted by $\tefflm$, $\logglm$, and $\fehlm$, while the bottom three subplots reveal the distribution of fitting residuals in relation to $\tefflm$, $\logglm$, and $\fehlm$, and the mean $\mu$ and standard deviation $\sigma$ of the fitting for each target star are provided. 

\begin{figure}
    \centering
    \includegraphics[width=1\linewidth]{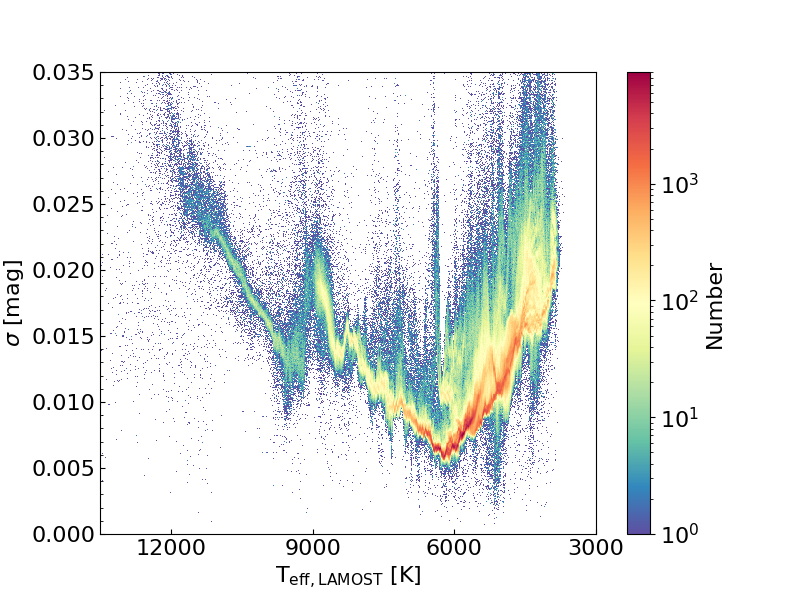}
    \caption{The 2D histogram of the residual standard deviation from the linear fit to the target stars, derived using the standard-pair algorithm, as a function of $\tefflm$, with color encoding representing the stellar number density.}
    \label{fig:star_pair_fit_sigma}
\end{figure}

\begin{figure*}
    \centering
    \includegraphics[width=1\linewidth]{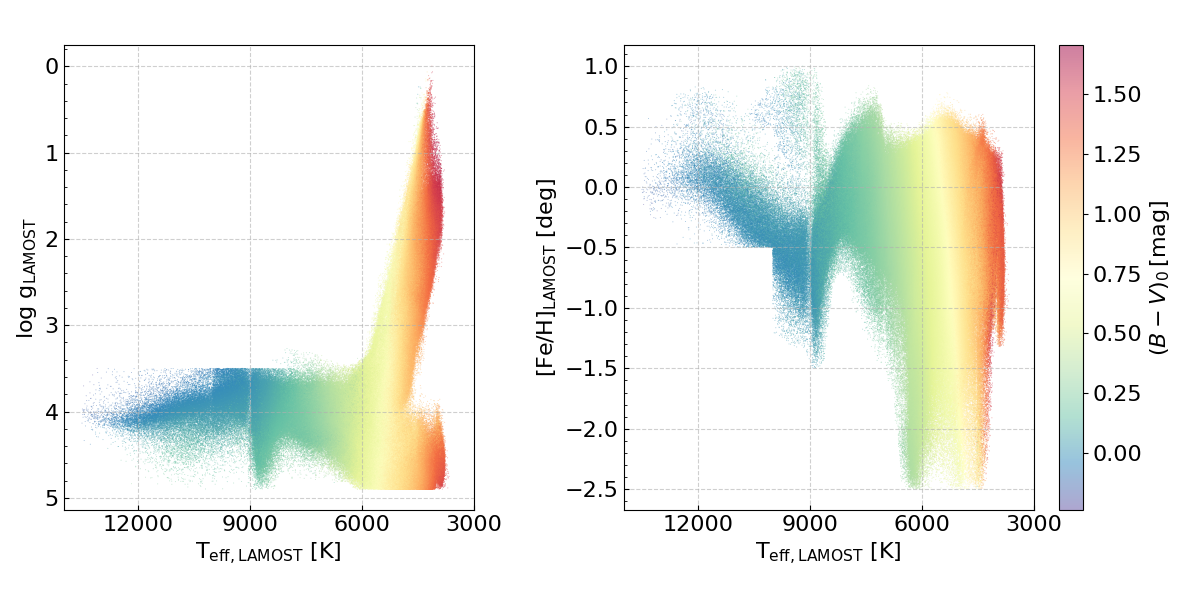}
    \caption{Distribution of intrinsic B$-$V colors in the stellar parameter space derived using the standard-pair algorithm. The color-coded dots represent the intrinsic B$-$V colors obtained from the LAMOST DR11 through the standard-pair algorithm.}
    \label{fig:Intrinsic_color_distribution}
\end{figure*}

To assess the fitting result of the overall sample, a 2D histogram of the distribution of the fitting standard deviation $\sigma$ with $\tefflm$ is presented in Figure \ref{fig:star_pair_fit_sigma}. The $\sigma$ here reflects the dispersion of the parameters. The discontinuity around 9000 K is due to a truncation in the stellar parameters from LAMOST at this temperature (see Figure \ref{fig:Intrinsic_color_distribution}). In contrast, the parameter measurements for G-type and F-type stars are the most precise, with the lowest fitting $\sigma$. Figure \ref{fig:star_pair_fit_sigma} also shows several upward 'spikes' in $\tefflm$ between 4000 and 7000 K, which correspond to metal-poor stars. This is because metal-poor stars are usually found in the more distant halo populations, are fainter and of lower SNR.

Using the 'standard-pair' algorithm, we have derived the intrinsic colors for $\sim$ 5 million stars. Figure \ref{fig:Intrinsic_color_distribution} illustrates the distribution of intrinsic colors within the stellar parameter space for the target sample. Generally, the intrinsic color varies smoothly with changes in stellar parameters.

For \ebv$_{\rm LAMOST}$, the reddening is calculated by \ebv$_{\rm LAMOST}$ = (B$-$V)$_{\rm observed}$ $-$ (B$-$V)$_{\rm intrinsic}$, where \ebv$_{\rm observed}$ is the observed color obtained from the $\gaia$ DR3 Synthetic Photometry catalog and \ebv$_{\rm intrinsic}$ is the intrinsic color predicted by the standard-pair algorithm.

\begin{figure}
    \centering
    \includegraphics[width=1\linewidth]{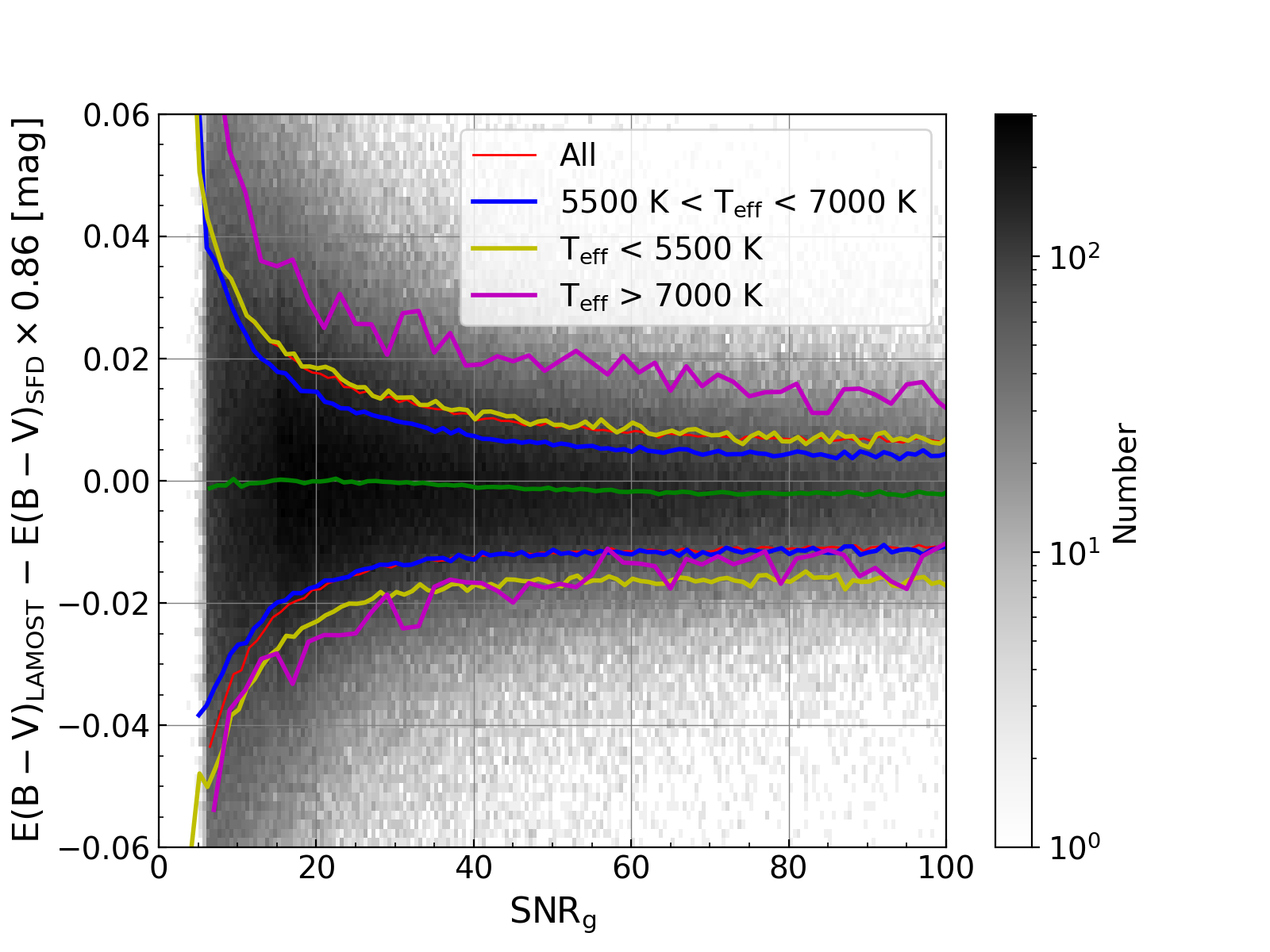}
    \caption{The 2D histogram of the difference between the algorithm-derived \ebv$_{\rm LAMOST}$ and $\sfdebv$ as a function of SNR$_g$, with color encoding representing the stellar number density. The data selection is based on \ebv < 0.02 mag , $|b| > 20^\circ $, and |Z| > 300 pc. Note that only stars with SNR$_g$ < 100 are shown. The red and green lines indicate the standard deviation $\sigma$ and $\mu$ computed within the bins of SNR$_g$, respectively. The blue, yellow-green, and magenta lines represent samples with $\tefflm$ in the ranges of 5500 K < $\tefflm$ < 7000 K, $\tefflm$ < 5500 K, and $\tefflm$ > 7000 K, respectively. The standard deviation $\sigma$ is calculated by binning the samples based on their SNR$_g$.}
    \label{fig:Diff_lm_sfd_vs_SNRg}
\end{figure}

To evaluate the precision of the obtained \ebv$_{\rm LAMOST}$, we selected stars from the final catalog with $\sfdebv$ < 0.02 mag, Galactic latitude |b| > 20$^\circ$, and a vertical distance from the Galactic plane |Z| > 300 pc, and compared them with the corresponding $\sfdebv$ derived from the SFD map. The above thresholds were chosen empirically to balance sample size and reliability for this comparison.
The variation in the differences between \ebv$_{\rm LAMOST}$ and $\sfdebv$ as a function of $g$-band SNR (SNR$_g$) is illustrated in Figure \ref{fig:Diff_lm_sfd_vs_SNRg}. For each SNR$_g$ bin, we calculated the mean $\mu$ and standard deviation $\sigma$ within the bin. 

Considering that the precision is not only related to the SNR$_g$ but also depends on stellar parameters, in Figure \ref{fig:Diff_lm_sfd_vs_SNRg}, we divide the sample into three temperature bins. For each bin, we calculate the standard deviation $\sigma$ by binning according to the SNR$_g$. Based on the trend of $\sigma$ with SNR$_g$, we smoothed the curve to assign a specific $\sigma$ to stars at each SNR$g$, representing the precision of their \ebv$_{\rm LAMOST}$ measurements. The mean extinction precision, \ebv$_{\rm err}$, is consistent with the result shown in Figure \ref{fig:star_pair_fit_sigma}, with the best extinction precision occurring in the temperature range of 5500 K - 7000 K. The median of \ebv$_{\rm err}$ for all approximately 4.6 million LAMOST sources obtained using this method is $\sim$ 0.014 mag.

The columns included in the final sample catalog are listed in Table \ref{table:lamost_ebv_catalog}. This catalog can be accessed via the public link: \url{https://nadc.china-vo.org/res/r101617/}.

\setlength{\tabcolsep}{1.5mm}{
\begin{table}[htbp]
\footnotesize
\centering
\caption{Description of the LAMOST catalog}
\begin{tabular}{ccc}
\hline
\hline
Field & Description & Unit\\
\hline
obsid & Unique spectra identifier for LAMOST DR11 & \\
source\_id & Unique source identifier for $\gaia$ DR3 & \\
ra &  R.A. from Gaia DR3 & deg\\
dec & Decl. from Gaia DR3 & deg\\
EBV & \ebv~ from this work & mag\\
EBV\_err & Error of \ebv~ from this work & mag\\
\hline
\end{tabular}
\label{table:lamost_ebv_catalog}
\end{table}}

\subsection{$\gaia$ DR3 data}\label{sec:Gaia}

The European Space Agency's (ESA) $\gaia$ mission is chiefly focused on acquiring astrometric, photometric, and spectroscopic data for celestial objects within the Milky Way and its Local Group \citep{2016AA...595A...1G}. In its third data release ($\gaia$ DR3), the mission has cataloged photometric information for more than 1.8 billion sources and has provided five-parameter astrometric solutions which include two positional coordinates, parallax, and two proper motion components for approximately 1.47 billion sources \citep{2023AA...674A...1G}. Additionally, $\gaia$ DR3 provides approximately 220 million low-resolution spectra, which are divided into those from the ‘Blue Photometer’ (BP) covering the wavelength range from 330 to 680 nm, and those from the ‘Red Photometer’ (RP) covering the range from 640 to 1050 nm \citep{2023AA...674A...1G,2023AA...674A...2D}. The parallaxes from $\gaia$, combined with the BP/RP spectra (hereafter, ‘XP spectra’), provide high-quality extinction and distance information.

\cite{zhang2023} developed a data-driven low-resolution Gaia DR3 BP/RP spectral model, using stellar atmospheric parameters from LAMOST as the training set, and incorporating photometric data from 2MASS and WISE. This model is used to infer the parameters (effective temperature $\teff$, surface gravity $\logg$, and $\feh$) as well as the corrected parallaxes and E(B$-$V) for approximately 220 million stars. In this study, we utilized the stellar parameter catalog from \cite{zhang2023}, referred to as the \href{https://doi.org/10.5281/zenodo.7692680}{Z23} \citep{zhang_2023_data}.

To construct a reliable three-dimensional extinction map of the Milky Way, we need to filter out reliable extinction measurements from the \href{https://doi.org/10.5281/zenodo.7692680}{Z23}. Based on the comparison with the extinction of co-sources from LAMOST and the "basic reliability" cut recommended in \cite{zhang2023}, we ultimately established the following cut criteria: \texttt{quality flag} < 8, \texttt{teff\_confidence} > 0.2, BP band SNR (SNR$_{\rm BP}$) > 5, and RP band SNR (SNR$_{\rm RP}$) > 7. 
Given that this study focuses on the extinction distribution within the MW, We excluded sources originating from the Large Magellanic Cloud (LMC) and the Small Magellanic Cloud (SMC) using the criteria provided by \cite{Huang2025a} (refer their Equation 6). Following the aforementioned selection process, we obtained a catalog comprising approximately 150 million sources with position and extinction information. To assess the extinction precision of this catalog, we cross-matched it with the LAMOST extinction data presented in subsection \ref{sec:Standard-Pair}, thereby identifying common sources for precision evaluation.

\begin{figure}
    \centering
    \includegraphics[width=1\linewidth]{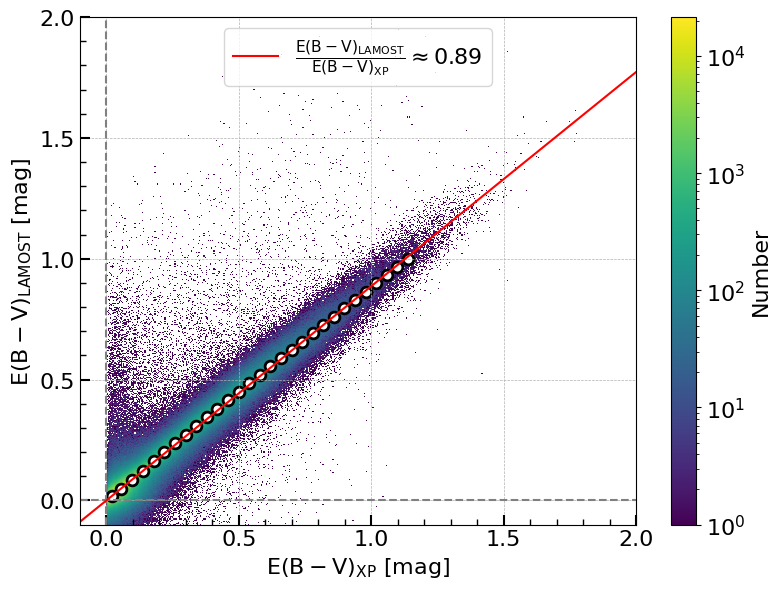}
    \caption{Comparison of \ebv derived from LAMSOT and XP. The hollow points represent the median within each bin, and the red line is the linear fit to the hollow points.}
    \label{fig:LAMSOT_vs_XP}
\end{figure}

We binned the common sources and calculated the median values for each bin. A linear fit to these median values yielded a ratio of approximately 0.89 (see Figure \ref{fig:LAMSOT_vs_XP}). To align the \ebv$_{\rm XP}$ with those from \ebv$_{\rm LAMOST}$, we adjusted the \ebv$_{\rm XP}$ by multiplying them by a factor of 0.89, which has been applied to all \ebv$_{\rm XP}$ throughout the remainder of this work. We computed the histogram distribution of the differences between \ebv$_{\rm LAMOST}$ and \ebv$_{\rm XP}$ and found that it follows a normal distribution with a $\sigma$ of $\sim$ 0.019 mag. Considering that \ebv$_{\rm LAMOST}$ has a dispersion of $\sim$ 0.011 mag, this indicates that for sources common to LAMOST and XP, the extinction precision of XP is $\sim$ 0.015 mag. At the same time, we also compared the sources outside the Galactic disk with $\sfdebv$ < 0.02 mag, and obtained similar precision.

\begin{figure}
    \centering
    \includegraphics[width=1\linewidth]{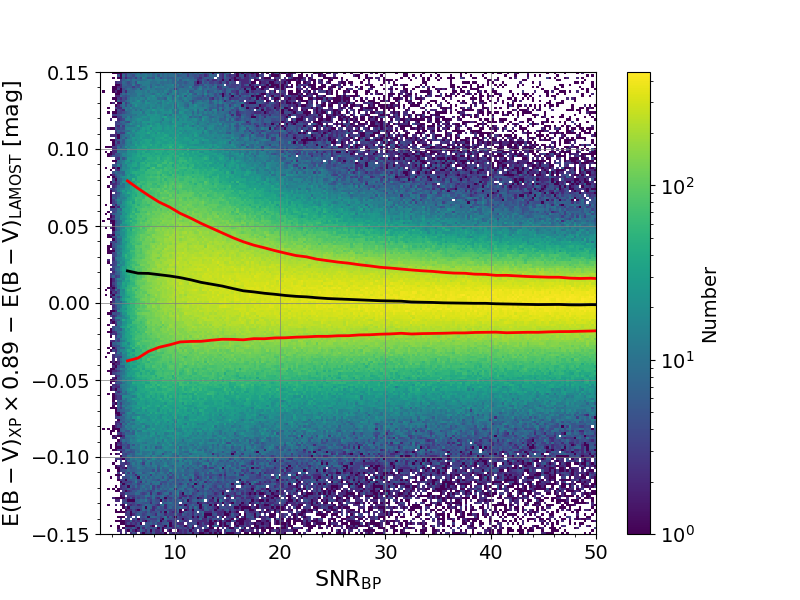}
    \caption{The 2D histogram of the difference between \ebv$_{\rm LAMOST}$ and \ebv$_{\rm XP}$ as a function of SNR$_{\rm BP}$, with color encoding representing the stellar number density. The black line represents the median value in each bin, while the red lines indicate the $\pm 1\,\sigma$ range.}
    \label{fig:XP_LM_sigma}
\end{figure}

We then plotted the difference between LAMOST and XP extinction as a function of the SNR$_{\rm BP}$ (see Figure \ref{fig:XP_LM_sigma}). For each bin, we calculated the dispersion, denoted as $\sigma$. The precision of the XP extinctions within each bin, $\sigma_{\rm XP} = \sqrt{\sigma^2 -\sigma^2_{\rm LAMOST}}$. By this method, we estimate the precision of XP extinction measurement $\sigma_{\rm XP}$.

It is worth noting from Figure \ref{fig:XP_LM_sigma} that a median offset of approximately 0.02 mag exists at low SNR. This offset mainly stems from the non-negativity constraint on extinction imposed by \cite{zhang2023}, which causes the random errors of low-extinction sources to scatter only toward higher values, thereby introducing a systematic median offset in the extinction estimates. Such an offset could systematically bias the inferred scale height of the dust disk toward higher values in high Galactic latitude regions. However, since these sources fall into the high-extinction-error category, they are typically assigned lower weights in subsequent fitting processes, thereby partially compensating for the aforementioned systematic offset.

We then compared the extinction precision $\sigma_{\rm XP}$ derived from our LAMOST data comparison with the \ebv$_{\rm err, Z23}$ provided in the \href{https://doi.org/10.5281/zenodo.7692680}{Z23}, plotting both against the SNR$_{\rm BP}$. We find that \ebv$_{\rm err, Z23}$ matches well with our derived $\sigma_{\rm XP}$. Therefore, we directly adopt \ebv$_{\rm err, Z23}$ as \ebv$_{\rm err}$ in the subsequent fitting procedure. The median value of \ebv$_{\rm err}$ in the filtered Z23 catalog is approximately 0.03 mag.

\subsection{Final data set}\label{sec:data_set}

Given that we consider the extinction precision from LAMOST to be superior to that from \href{https://doi.org/10.5281/zenodo.7692680}{Z23}, for the common sources, we opted to retain the extinction from LAMOST. For the extinction in Z23, we applied a multiplicative factor (see Section \ref{sec:Gaia}) to align them with the LAMOST extinction.

Since the $\gaia$ XP data do not cover the full extent of the LAMOST dataset, for sources unique to LAMOST, we adopted the distance information from \cite{BJ2021}. Specifically, we used \texttt{rpgeo} as the distance estimate, with \texttt{B\_rpgeo} and \texttt{b\_rpgeo} representing the upper and lower distance uncertainties, respectively. For the common sources and those from \href{https://doi.org/10.5281/zenodo.7692680}{Z23}, we utilize the revised parallax and parallax uncertainty provided by \href{https://doi.org/10.5281/zenodo.7692680}{Z23}. The distance is simply calculated as the reciprocal of the parallax, and the distance uncertainty is derived as $\frac{\text{parallax uncertainty}}{\text{parallax}} \times \text{distance}$.

This resulted in a dataset of approximately 150 million sources, each with positional information, \ebv, \ebv$_{\rm err}$, distances, and distance uncertainties. This comprehensive dataset will serve as the foundation for constructing the 3D map.

\section{Method} \label{sec-Method}

\subsection{Adaptive multi-resolution line-of-sight}\label{sec:line-of-sight}

To analyze the characteristics of the data density distribution in Galactic coordinates, we employed the Hierarchical Equal Area isoLatitude Pixelization (HEALPix) scheme \citep{HEALPix}. This method divides the entire sphere into a series of equal-area, approximately diamond-shaped pixels. The number of pixels and their respective areas are determined by a single parameter, $\rm N_{side}$, which is defined as a power of 2 (i.e., $\rm N_{side}$ = 2$^k$, where $k$ = 0, 1, 2, ...).  This parameter divides the entire sphere into 12$\times \rm N_{side}^2$ equal-area pixels.

\begin{figure*}
    \centering
    \includegraphics[width=1\linewidth]{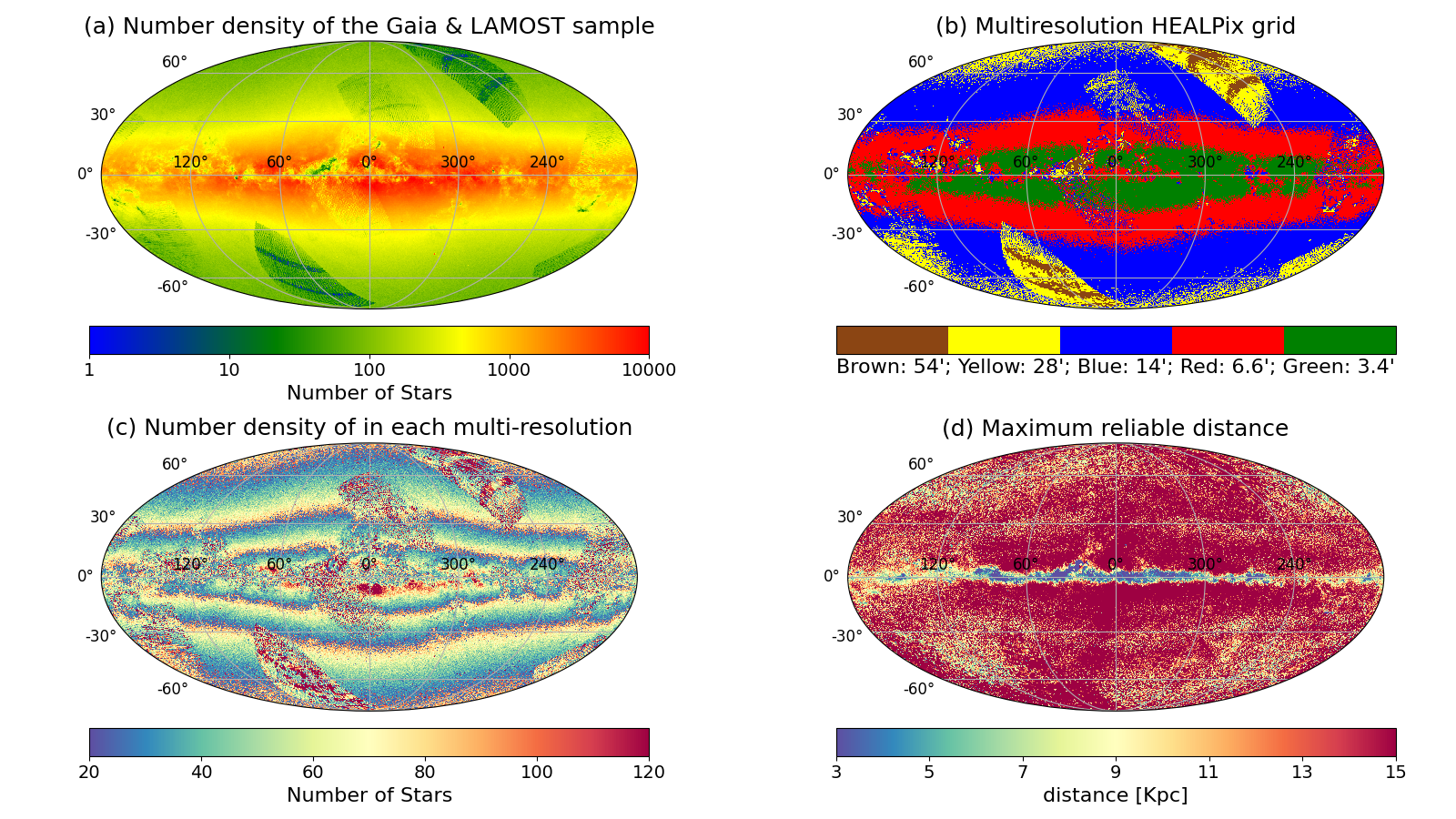}
    \caption{Adaptive multi-resolution line-of-sight scheme maximum reliable distance for each line-of-sight. (a) The number density of the LAMOST and XP sample（see subsection \ref{sec:data_set}）in the Galactic coordinate system. The number of stars per $\sim$ 0.21 deg$^2$ area（$\rm N_{side}$ =128）is shown. (b) A adaptive multi-resolution map created using the data shown in panel (a), in the same coordinate system. The map consists of five distinct HEALPix pixels, each represented by a different color: brown represents a resolution of 54\arcmin (area $\sim$ 0.84$\ \mathrm{deg}^2$), yellow represents 28\arcmin (area $\sim$ 0.21$\ \mathrm{deg}^2$), blue represents 14\arcmin (area $\sim$ 0.05$\ \mathrm{deg}^2$), red represents 6.6\arcmin (area $\sim$ 0.01$\ \mathrm{deg}^2$), and green represents 3.4\arcmin (area $\sim$ 0.003$\ \mathrm{deg}^2$). (c) The distribution of the number of stars per line-of-sight, derived from the adaptive multi-resolution scheme shown in panel (b), is color-coded according to the stellar count. (d) In the adaptive multi-resolution map, the maximum reliable distance for each line-of-sight is color-coded.}
    \label{fig:HEALPix}
\end{figure*}

In Figure \ref{fig:HEALPix}(a), we selected $\rm N_{side} = 128$, which corresponds to a pixel area of approximately 0.2 deg$^2$. In the figure, the number of stars within each pixel is represented through color coding, which demonstrates the stellar number density distribution. The arc-like streaks visible in the figure are caused by the Gaia scanning law. Furthermore, the pronounced non-uniformity of the number density along the Galactic plane is primarily due to significant extinction effects, which are a result of the substantial and uneven distribution of foreground dust.

To address the uneven distribution of stellar density, we adopted a multi-resolution map scheme based on HEALPix, where the pixel size dynamically varies with the observed stellar density. In HEALPix, each pixel can be subdivided into four equal-area smaller pixels at a higher resolution (by increasing the power of $k$ in $\rm N_{side}$). This allows us to dynamically adjust the pixel area in two ways:

1. Top-down subdivision scheme: The entire sphere is initially divided into larger resolution pixels. If the number of stars within a pixel exceeds a predefined threshold, the pixel is recursively subdivided into four smaller pixels until the number of stars in each pixel is below the threshold. However, the drawback of this method is that, in regions with large variations in stellar distribution, some subpixels may contain significantly fewer stars than one-quarter of the threshold after subdivision;

2. Bottom-up subdivision scheme: The sphere is first divided into sufficiently small pixels, and then the number of stars in each pixel is examined. If the number of stars in a pixel is below the predefined threshold, the pixel is merged with its three neighboring pixels belonging to the same higher-level $\rm N_{side}$, forming a pixel at the higher-level $\rm N_{side}$. This process continues until the number of stars in each merged pixel exceeds the threshold. This approach effectively adjusts the size of $\rm N_{side}$ pixels, ensuring that the number of stars in each pixel is not less than the threshold.

To ensure that each pixel contains a sufficient number of stars for subsequent reddening modeling, we adopted the second bottom-up subdivision strategy. Since the following analysis allows for considering information beyond individual pixels, we subdivided the pixels more aggressively. We set a threshold of at least 20 stars per pixel.

Based on this subdivision criterion, we ultimately obtained five different resolutions of HEALPix pixel divisions, namely: $\rm N_{side}$ = 64 ($\sim$ 0.84 deg$^2$), $\rm N_{side}$ = 128 ($\sim$ 0.21 deg$^2$), $\rm N_{side}$ = 256 ($\sim$ 0.05 deg$^2$), $\rm N_{side}$ = 512 ($\sim$ 0.01 deg$^2$), and $\rm N_{side}$ = 1024 ($\sim$ 0.003 deg$^2$). The positions of the pixels with different resolutions on the sky are shown in Figure \ref{fig:HEALPix}(b), and the specific information is provided in Table \ref{tab:pix}. The number of stars contained in each multi-resolution pixel is also shown in Figure  \ref{fig:HEALPix}(c), with a median value of 43 stars (See Table \ref{tab:pix}).

\setlength{\tabcolsep}{1.5mm}{
\begin{table*}[htbp]
\footnotesize
\centering
\caption{Pixelization of the sky}
\begin{tabular}{ccccccc}
\hline
\hline
$\rm N_{side}$ & pixel scale& Pixel area ($\mathrm{deg}^{2}$)& $\Omega$ ($\mathrm{deg}^{2}$)&Number of pixels & proportion & The median of stars in pixels\\
\hline
64    &  $\sim 55^{\prime}$   &  $\sim$ 0.84   &  1077.65     &  1284      &  3\%  & 127\\
128   &  $\sim 27^{\prime}$   &  $\sim$ 0.21   &  4781.24     &  22787    &  12\% & 86\\
256   &  $\sim 14^{\prime}$   &  $\sim$ 0.05   &  17104.96    &  326083    &  41\% & 48\\
512   &  $\sim 6.9^{\prime}$  &  $\sim$ 0.01   &  12554.48  &  957337  &  30\% & 47\\
1024   &  $\sim 3.4^{\prime}$  &  $\sim$ 0.003   &  5734.62  &  1749164   &  14\% &40\\
\hline
totals &  ---  &  ---  &  41252.96  &  2910405  & 100\% &43\\[2pt]
\end{tabular}
\label{tab:pix}
\end{table*}}

Meanwhile, we chose to involve the stars surrounding each line-of-sight pixel in the subsequent reddening modeling. Each star is assigned a weight, with stars closer to the given line-of-sight center being more representative, and the weight decreasing as the distance from the line-of-sight increases. We use the function $\omega_{i, \text{decentration}}$ to describe the assigned weight:
 
\begin{equation}
\omega_{i, \text{decentration}} = \frac{1}{1 + e^{(\alpha/A_r)\theta + \beta}}
\label{eq:weight_decentration}
\end{equation}
where $\theta$ is the angular distance between star $i$ and the selected line-of-sight center, $A_r$ is the angular resolution of the pixel, and \st{$a$ and $b$} $\alpha$ and $\beta$ are parameters used to adjust the weight distribution. 

\begin{figure}
    \centering
    \includegraphics[width=1\linewidth]{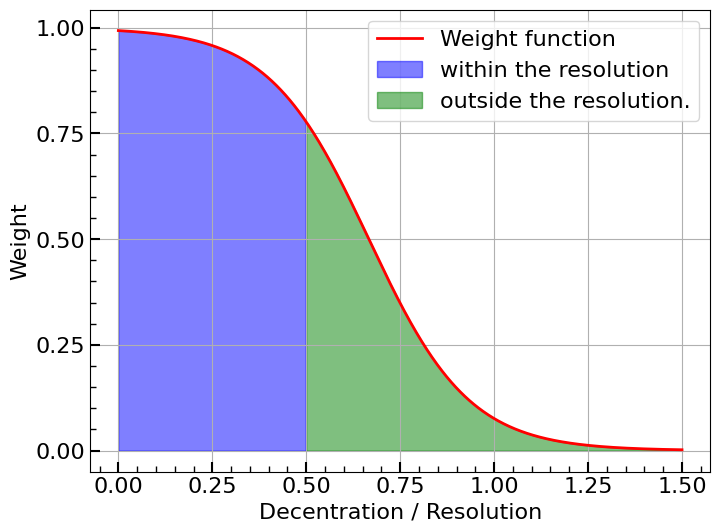}
    \caption{Weight function of stellar decentration relative to resolution. The horizontal axis represents the ratio of decentration to resolution, with both decentration and resolution measured in degrees. The blue region indicates decentration within the resolution limit, while the green region represents decentration beyond the resolution limit. The volume of the regions formed by rotating the blue and green areas around the y-axis is approximately equal, implying that, under the assumption of a roughly uniform stellar density, the contributions (weights) from the region within the resolution limit and the region beyond the resolution limit are approximately balanced.}
    \label{fig:Weight}
\end{figure}

To ensure that the resolution is not diluted by the influence of surrounding stars, we require that the total weight of all stars outside the resolution range should be equal to the total weight of the stars within the resolution range, implying that, under the assumption of a roughly uniform stellar density, the contributions (weights) from the region within the resolution limit and the region beyond are approximately balanced. Specifically, this is equivalent to ensuring that the volume of the solid of revolution of the blue area in Figure \ref{fig:Weight} is approximately equal to the volume of the solid of revolution of the green area. Under this condition, we obtained the parameters $\alpha = 7.5$ and $\beta = -5$.

Using this method, we select all stars within a circle of three times the angular resolution centered on each resolution pixel, and assign a weight $\omega_{i, \text{decentration}}$ to each star. These weighted stellar data will serve as the fundamental data source for subsequent reddening modeling along the line-of-sight for that pixel.

\subsection{The Extinction-distance Model}\label{sec:function}

In the absence of molecular clouds, interstellar extinction increases smoothly with distance, primarily due to the diffuse interstellar medium (DIM) of the MW. However, when molecular clouds are present, the extinction exhibits a significant jump at the distance corresponding to the location of the cloud. Therefore, the overall extinction within the Galaxy can be considered as a combination of a gradual increase due to the diffuse dust component, superimposed with local discontinuities caused by molecular clouds. It is worth mentioning that there are also many local cavities in the DIM, caused by supernova-driven processes (e.g., \citealt{1992ApJ...388...93K,1995ASPC...80..292M,2001RvMP...73.1031F}).

Thus, for any given line-of-sight, the total color excess E(B$-$V) is comprised of two components: the color excess due to the diffuse interstellar medium \ebv$^{\mathrm{DIM}}$, and the color excess due to multiple molecular clouds \ebv$^{\mathrm{MC}}$:

\begin{equation}
\mathrm{E(B-V)} = \mathrm{E(B-V)} ^{\mathrm{DIM}} + \sum_{i=1}^{n} \mathrm{E(B-V)} ^{\mathrm{MC}}_i
\end{equation}

For \ebv$^{\mathrm{DIM}}$, we assume that the DIM is the maximum density at the Galactic plane and that this density decreases exponentially with increasing vertical distance from the plane. Notably, the Sun is located in a low dust density region formed by supernovae activity, known as the Local Bubble (e.g., \citealt{1987ARA&A..25..303C,2006MNRAS.373..993F,2009Ap&SS.323....1W,2021ApJ...920...75L,2022Natur.601..334Z}), where the dust density is negligible. Therefore, we use a piecewise function to represent the contribution of DIM along the line-of-sight, with the derivative of this function exhibiting exponential decay,

\begin{equation}
\mathrm{E(B-V)} ^{\mathrm{DIM}} =
\begin{cases} 
0, & d \leq d_{\text{bubble}} \\
\frac{h}{\sin b^\circ} \times \rho_{\mathrm{max}} \times \left( 1 - e^{-\frac{d}{h / \sin b^\circ}} \right), & d > d_{\text{bubble}}
\end{cases}
\end{equation}
where, $d$ denotes the distance from the Sun (assuming the Sun lies on the Galactic plane), $d_{\rm bubble}$ denotes the distance from the Sun to the boundary of the Local Bubble. The Local Bubble is a cavity formed by supernova-driven processes within the previously established DIM structure. Consequently, the formation of the Local Bubble does not alter the overall distribution characteristics of the DIM. The dust that is closer to the supernova will be evaporated and destroyed, while the dominant dust is swept and compressed, transferring to the boundary of the local cavity \citep{2020A&A...639A..72W}. As a result, the color excess within the Local Bubble can be considered negligible until it abruptly returns to the typical DIM levels at the boundary. The parameter $h$ represents the DIM scale height in the line-of-sight, while $b$ is the Galactic latitude in the line-of-sight. Thus, $h/\sin b$ represents the projection of the scale height $h$ along the line-of-sight, which physically means that the DIM density will decrease to $1/e$ of its maximum value along the line-of-sight within this projected distance. $h/\sin b \times \rho_{\mathrm{max}}$ represents the cumulative color excess after passing through the entire DIM in the line-of-sight, where $\rho_{\mathrm{max}}$ denotes the DIM density at the location of the Sun in the Galactic plane in the absence of the Local Bubble, which is the maximum density along the vertical line from the Sun to the Galactic plane. Compared to directly using a free parameter to represent the cumulative color excess of DIM along the line-of-sight, this method provides a more effective way to constrain the cumulative DIM contribution, thereby allowing for a clearer distinction between the reddening effects of diffuse dust and those from nearby molecular clouds in the reddening model.

\begin{figure*}
    \centering
    \includegraphics[width=1\linewidth]{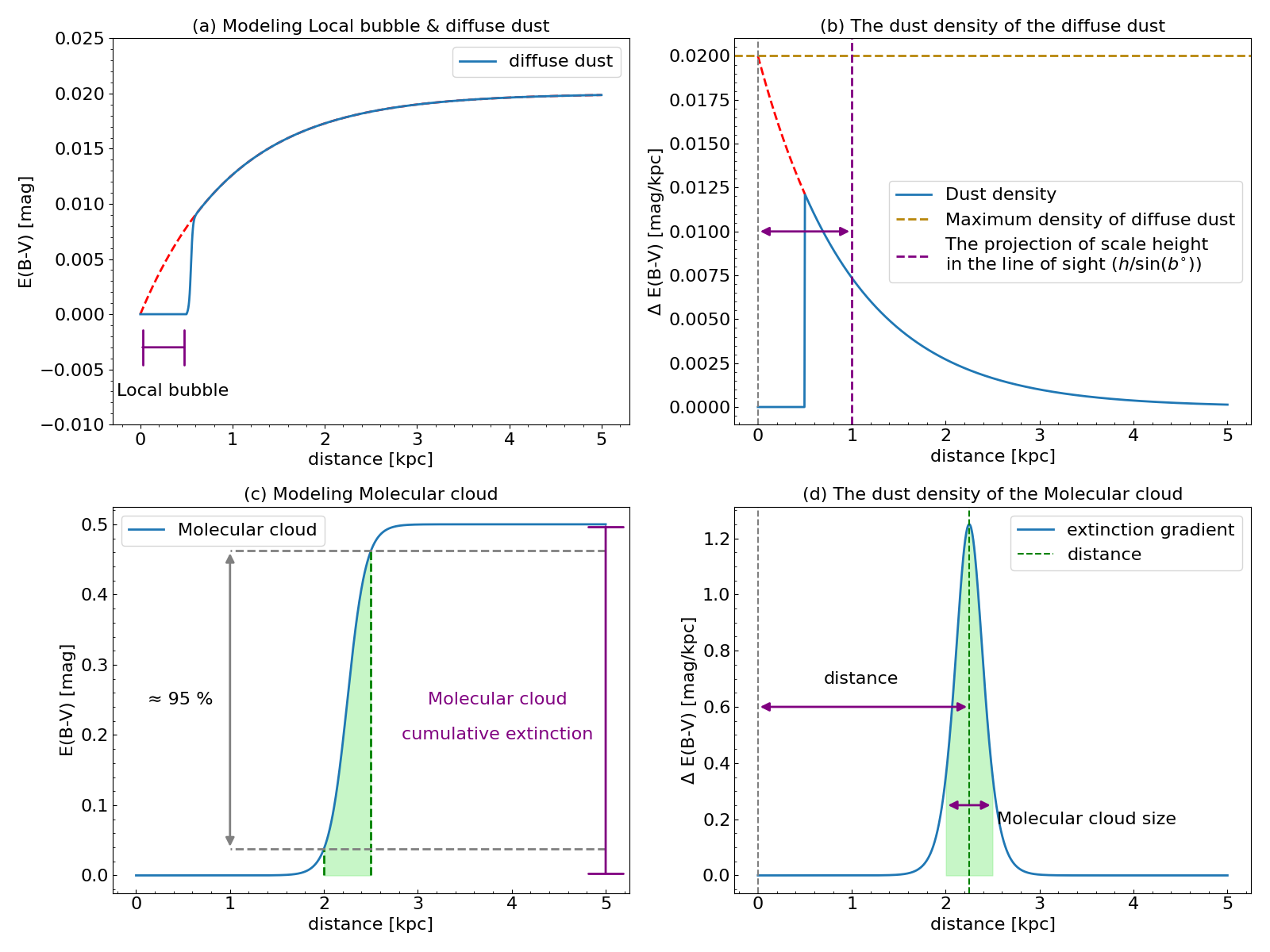}
    \caption{Modeling the extinction-distance relationship. The x-axis represents distance, while the left panels show extinction, and the right panels show the extinction gradient. (a) The blue line represents the extinction model for diffuse dust. The purple segment highlights one of the three free parameters used in the model: the distance from the Sun to the peak extinction surface of the local bubble. The red dashed line shows the extinction function without the local bubble. (b) The blue line shows the derivative of the extinction function. The red dashed line represents the extinction gradient in the absence of the local bubble, while the gold dashed line marks the extinction gradient value at the point where the distance equals zero. This gradient value corresponds to another free parameter used in the model: the maximum density of the diffuse dust. The purple segment indicates the distance at which the dust density falls to 1/e of its maximum value, which corresponds to the projection of the diffuse dust scale height along the line-of-sight, another key free parameter. (c) The blue line represents the extinction function for the molecular cloud model. The purple segment indicates the cumulative extinction through the molecular cloud, while the green shaded region corresponds to the distance at which approximately 95\% of the cumulative extinction is enclosed, defining the parameters related to the molecular cloud size, denoted as $\Lambda^{\mathrm{MC}}_i$. (d) The blue line shows the derivative of the molecular cloud extinction function. The green dashed line marks the distance where the extinction gradient peaks, defined as the molecular cloud distance. The green shaded region represents the molecular cloud size, as defined in panel (c), while the area beneath the blue line represents the cumulative extinction through the molecular cloud.}
    \label{fig:function}
\end{figure*}

The independent variable here is the distance along the line-of-sight, with the overall function depicted in the upper left panel of Figure \ref{fig:function}, and its derivative shown in the upper right panel. 

To model the potential multiple components of molecular clouds, we assume that a molecular cloud exhibits a structure with maximum density at its center, with the density decreasing exponentially toward the periphery. This configuration is effectively captured using a series of n modified sigmoid functions. These functions are particularly well-suited for representing variations in \ebv.  A single function is illustrated in the lower left panel of Figure \ref{fig:function}, with its equation as follows:

\begin{equation}
\begin{split}
\mathrm{E(B-V)}^{\mathrm{MC}}_i = (1 + a)\times \Delta \mathrm{E(B-V)}^{\mathrm{MC}}_i \\
\times \left(\frac{1}{1 + e^{ -5 \cdot \frac{d - d^{\mathrm{MC}}_i}{\Lambda^{\mathrm{MC}}_i} }}-b\right)
\end{split}
\end{equation}
where, $\Delta \mathrm{E(B-V)}^{\mathrm{MC}}_i$ represents the cumulative reddening along the entire line-of-sight through the molecular cloud. The parameter 
$d^{\mathrm{MC}}_i$ denotes the distance from the Sun to the center of the molecular cloud, while $\Lambda^{\mathrm{MC}}_i$ represents the extent of the cloud along the line-of-sight, defined as the distance over which $\sim$ 95\% of the total cumulative reddening occurs. Since our modeling begins at a distance of zero for each line-of-sight, it is necessary for the function to precisely pass through the origin. To achieve this, we introduce two constants, a and b, to shift the function, ensuring that it intersects the zero point. In most scenarios, a and b are close to zero. However, when the distance to the molecular cloud d is small and the size of the cloud is relatively large, the values of a and b may become significant. These constants are defined as follows:

\begin{equation}
a = \frac{1}{e^{5 \cdot d^{\mathrm{MC}}_i/\Lambda^{\mathrm{MC}}_i}}
\end{equation}

\begin{equation}
b = \frac{1}{1 + e^{ 5 \cdot d^{\mathrm{MC}}_i/\Lambda^{\mathrm{MC}}_i }}
\end{equation}

The function we have constructed is a combination of multiple modified elementary functions, which allows us to conveniently obtain an analytical expression for its derivative. Specifically, this derivative refers to $dE(B-V)/d\mathrm{Distance}$, that is, the rate of change of extinction with respect to distance along the line of sight.  Physically, it represents the density of dust extinction per unit distance. This quantity is expressed as:

\begin{equation}
\mathrm{E(B-V)}' = \left (\mathrm{E(B-V)}^{\mathrm{DIM}}\right ) ' + \sum_{i=1}^{n} \left (  \mathrm{E(B-V)}^{\mathrm{DIM}}_{i}\right) '
\end{equation}
where:

\begin{equation}
\left (\mathrm{E(B-V)}^{\mathrm{DIM}}\right ) ' =
\begin{cases} 
0,& d \leq d_{\text{bubble}}  \\
 \rho_{\mathrm{max}} \times e^{-\frac{d}{h / \sin b^\circ}}, & d > d_{\text{bubble}} 
\end{cases}
\end{equation}

\begin{equation}
\begin{split}
\left ( \mathrm{E(B-V)}^{\mathrm{MC}}_{i}\right ) ' = \frac{5 \times (1 + a) \times \Delta \mathrm{E(B-V)}^{\mathrm{MC}}_i}{\Lambda^{\mathrm{MC}}_i} \nonumber \\ \times \mathrm{S}(d)(1-\mathrm{S}(d))
\end{split}
\end{equation}

with:

\begin{equation}
\mathrm{S}(d) = \frac{1}{1 + e^{ -5 \cdot \frac{d - d^{\mathrm{MC}}_i}{\Lambda^{\mathrm{MC}}_i} }}
\end{equation}

\subsection{The Model Fitting} \label{sec:fitting}

In each line-of-sight, it is essential not only to account for the weight variations arising from the angular distance from the line-of-sight center but also to simultaneously consider the influences of distance uncertainties and \ebv uncertainties. These combined factors contribute to the overall weight assigned to each data point, ensuring a more accurate representation of their reliability in the analysis.

The uncertainty in reddening is represented by \ebv$_{\mathrm{err}}$. Initially, a preliminary weight factor was calculated as follows:

\begin{equation}
\omega_{\mathrm{E(B-V)}_{\mathrm{err}}} = 1 / \mathrm{E(B-V)}_{\mathrm{err}}^ 2,
\end{equation}

To prevent any data point from disproportionately influencing the results due to extremely high or low uncertainties, upper and lower limits were imposed on the weight values. Specifically, the maximum weight was set to correspond to an \ebv${\mathrm{err}}$ of 0.01 mag, while the minimum weight was set to correspond to an \ebv${\mathrm{err}}$ of 0.1 mag. It is worth noting that 99.99\% of our sample have \ebv$_{\mathrm{err}}$ smaller than 0.1 mag, so the minimum weight only affects a negligible fraction of the data. Finally, the resulting weight factors were normalized to range between 10\% and 100\%.

\begin{equation}
\omega_{\mathrm{E(B-V)}_{\mathrm{err}}}^{\rm norm} = \left( \frac{\omega_{\mathrm{E(B-V)}_{\mathrm{err}}} - \frac{1}{0.1^2}}{ \frac{1}{0.01^2} - \frac{1}{0.1^2} } \right) \times (1 - 0.1) + 0.1
\end{equation}

For the distance uncertainty, we simply calculate it as:$d_{\rm err}^{\rm absolute} = (d_{\rm err}^{\rm up} +  d_{\rm err}^{\rm low}) / 2$, where $d_{\rm err}^{\rm up}$ and $d_{\rm err}^{\rm low}$ are respectively the upper and lower bounds of the 68 per cent confidence interval of the distance estimate.

Since the uncertainty in \ebv$_{\mathrm{err}}$ is expressed in magnitudes, it inherently represents a relative measurement.  In contrast, the distance is given in kpc, which is an absolute measure.  Therefore, before assigning weights to the distance, it is necessary to convert the error into a relative uncertainty: $d_{\rm err}^{\rm relative} = d_{\rm err}^{\rm absolute} / d$.

We opted to use the reciprocal of the relative uncertainty, rather than its square, to calculate an initial weight factor:

\begin{equation}
\rm \omega_{distance_{\mathrm{err}}} = 1 / \text{distance}_{\rm err}^{\rm relative},
\end{equation}

This choice is motivated by the fact that using the squared reciprocal could lead to an extreme distribution of weights. Specifically, many data points with larger $d_{\mathrm{err}}^{\rm relative}$ might end up with weights close to zero, significantly diminishing their influence in the fitting process and potentially causing the model to disregard these data points entirely. By using the reciprocal of the $d_{\mathrm{err}}^{\rm relative}$, rather than its square, we ensure that the weights maintain a reasonable level of differentiation, allowing all data points to contribute meaningfully to the analysis. Similarly, as with the uncertainty in reddening, the weights based on relative distance uncertainties were also constrained to avoid extreme effects. Specifically, the maximum weight was set to correspond to a $d_{\mathrm{err}}^{\rm relative}$ of 0.5\%, while the minimum weight was set to correspond to a $d_{\mathrm{err}}^{\rm relative}$ of 50\%. The resulting weight factors were then normalized to a range between 10\% and 100\%, ensuring consistency in subsequent analyses：

\begin{equation}
\rm \omega_{d_{\mathrm{err}}}^{norm} = \left( \frac{\omega_{d_{\mathrm{err}}} - \frac{1}{0.5^2}}{ \frac{1}{0.005^2} - \frac{1}{0.5^2} } \right) \times (1 - 0.1) + 0.1
\end{equation}

The weights derived from the uncertainties in \ebv~ and distance were combined with the weights calculated in Section \ref{sec:line-of-sight}, which were based on the angular distance from the line-of-sight center, to compute the final weight for each data point:

\begin{equation}
\rm \omega_{total} = \omega^{decentration} \times \omega_{\mathrm{E(B-V)}_{\mathrm{err}}}^{norm} \times \omega_{distance_{\mathrm{err}}}^{norm}
\end{equation}

The number of molecular clouds along any given line-of-sight is unknown. Therefore, in equation (1), we initially choose four components (i.e., n=4). In most cases, four molecular clouds are sufficient to cover the possible clouds along a line-of-sight, as more clouds would result in significant extinction, preventing us from observing the background stars. Thus, four components are adequate. Our primary focus is to identify the truly significant components while minimizing the influence of redundant ones. To achieve this, we constructed the following loss function:

\begin{align}
L(\rm param) = \sum_{j=1}^N (\omega_{\rm total})_j \left| \mathrm{E(B-V)}_j - \hat{E}(B-V)_j \right| \nonumber \\ + \lambda \sum_{i=1}^4 \left| \Delta \mathrm{E(B-V)}^{\mathrm{MC}}_i \right|
\end{align}

Here, $j$ indexes individual stars along a given line of sight, and $N$ denotes the total number of stars in that sightline. The loss function consists of two components: a weighted loss and a regularization term.

The first component is the weighted loss, which measures the difference between the predicted value and the actual observed value, adjusted by the weight coefficient $(\omega_{\rm total})_j$. We use the absolute value of the residual, $\left| \mathrm{E(B-V)}_j - \hat{E}(B-V)_j \right|$, instead of the squared residual to construct the weighted loss, as the absolute value is less sensitive to outliers, making the model more robust and better to handle anomalies caused by measurement errors or circumstellar dust. In contrast, the squared residual is overly sensitive to outliers, resulting in larger loss values due to the magnification of larger deviations, thereby causing the model to focus excessively on the influence of outliers.

The second component is the regularization term, which aims to select reliable components and reduce redundancy. Here, we use the sum of the absolute values of the cumulative reddening for each component ($\Delta \mathrm{E(B-V)}^{\mathrm{MC}}_i$), multiplied by the regularization coefficient $\lambda$, to form the L1 regularization penalty. L1 regularization (sum of absolute values) effectively removes unnecessary components by driving some model parameters to zero, resulting in a sparse solution. In contrast, L2 regularization (sum of squares) tends to smooth all parameters and cannot completely remove redundant features. To select features, we choose L1 regularization, and the regularization strength is adjusted by the coefficient $\lambda$. A larger $\lambda$ increases the tendency to eliminate features. For the $\lambda$, we set it to 0.1.

To solve the model parameters, we used the \texttt{scipy.optimize.minimize()} function to minimize the custom loss function. Specifically, we adopted the L-BFGS-B (Limited-memory Broyden-Fletcher-Goldfarb-Shanno with Bounds) algorithm for optimization. L-BFGS-B is a limited-memory quasi-Newton method that is particularly suitable for optimization problems with boundary constraints, which allows us to restrict the model parameters within reasonable physical ranges.

\setlength{\tabcolsep}{1.5mm}{
\begin{table*}[htbp]
\footnotesize
\centering
\caption{The unit, initial, lower and upper boundary of the parameter.}
\begin{tabular}{ccccccc}
\hline
Parameters & Unit& Initial & Lower boundary& Upper boundary \\
\hline
$d_{\rm bubble}$            & pc      &100 &50 & 1000  \\
$h$                         & pc      &120 &50  & 500 \\
$\rho_{\mathrm{max}}$                & mag/kpc &0.02&0 & 0.04 \\
$d^{\mathrm{MC}}_i$       & kpc     &$d_i$\tablenotemark{a}       & $ d_{\rm bubble}$  & 15 \\
$\Lambda^{\mathrm{MC}}_i$           & kpc     & 0.1 $\times d_i$& 0.01 & 0.5 $\times d^{\mathrm{MC}}_i$ \\
$\Delta \mathrm{E(B-V)}^{\mathrm{MC}}_i$  & mag     & $\rm E(B-V)_{max}/4$\tablenotemark{b}   & 0 & 3 \\
\hline
\end{tabular}
\tablenotetext{a}{$d_i$ comprises four sets of distinct initial values, with each set containing four different distances used for locating molecular clouds at various positions.}
\tablenotetext{b}{$\rm E(B-V)_{max}$ represents the approximate maximum extinction for stars in this line-of-sight, determined after applying a 3-$\sigma$ culling criterion to the data.}
\label{tab:param}
\end{table*}}

Our model includes three parameters related to the diffuse dust and three parameters for each molecular cloud. For the optimization process, initial values and boundaries for the model parameters are provided (see Table \ref{tab:param}). It is important to note that due to the significant variations in the distances of molecular clouds, a common challenge faced during the optimization is that the algorithm tends to become trapped in local minima, failing to reach the global optimum. To overcome this limitation and increase the probability of finding the global optimum, we employ a multi-start strategy, conducting multiple independent fitting attempts. Specifically, we prepare four sets of different initial parameters of molecular cloud distance (varying only the molecular cloud parameters), with each set representing a separate fitting experiment. This approach aims to explore different regions of the solution space, thereby enhancing the likelihood of identifying the global optimum.

For each line-of-sight, we conducted multiple fitting attempts and evaluated the results using the standard deviation of the residuals ($\sigma$) as our assessment metric. The set of parameters yielding the minimum $\sigma$ was selected, as it provides the closest approximation to the global optimum under the current conditions.  However, some data points exhibit significant deviations from the fitted curve, likely due to the influence of circumstellar dust or measurement errors.  To address these outliers, we implemented an iterative removal process, whereby data points deviating by more than 3-$\sigma$ were excluded. Subsequently, using the optimal parameters obtained from the initial fitting as new starting values, we performed a refitting of the dataset after the exclusion of outliers. This process ultimately yielded more precise fitting parameters, which we adopted as our final results.

\section{Result} \label{sec-Result}
\subsection{3D Reddening Map}\label{sec:map}

\begin{figure}
    \centering
    \includegraphics[width=1\linewidth]{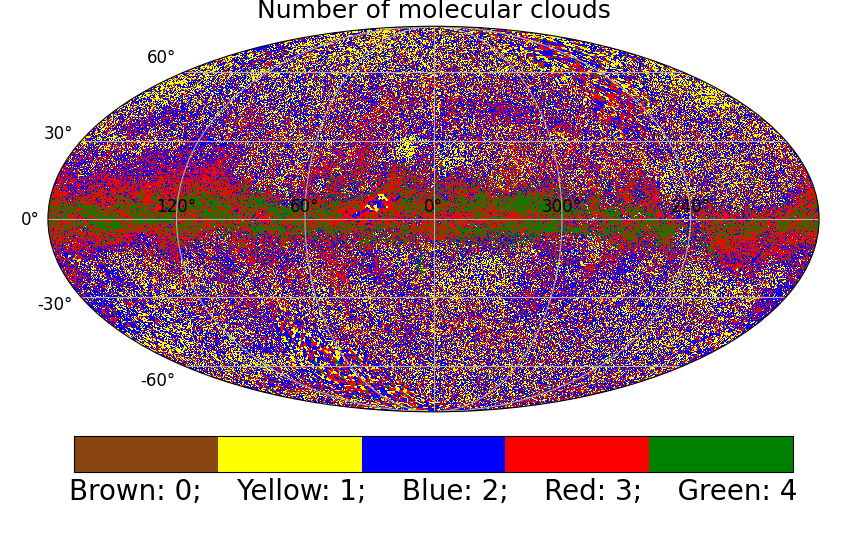}
    \caption{The number of independent molecular clouds along the line-of-sight is considered. The brown, yellow, blue, red, and green regions represent the presence of 0, 1, 2, 3, and 4 molecular clouds, respectively.}
    \label{fig:hp_comp}
\end{figure}

For each line-of-sight, we obtained a set of parameters that allowed us to construct continuous distance-reddening curves, which shows examples of distance-reddening curves for multiple lines-of-sight with overlaid data. In subsequent analyses, we excluded components with cumulative extinction parameters ($\Delta \mathrm{E(B-V)}^{\mathrm{MC}}_i$) less than 0.001 mag. Consequently, each line-of-sight includes contributions from diffuse dust as well as 0 to 4 distinct significant components (see Figure \ref{fig:hp_comp}).

\begin{figure*}
    \centering
    \includegraphics[width=1\linewidth]{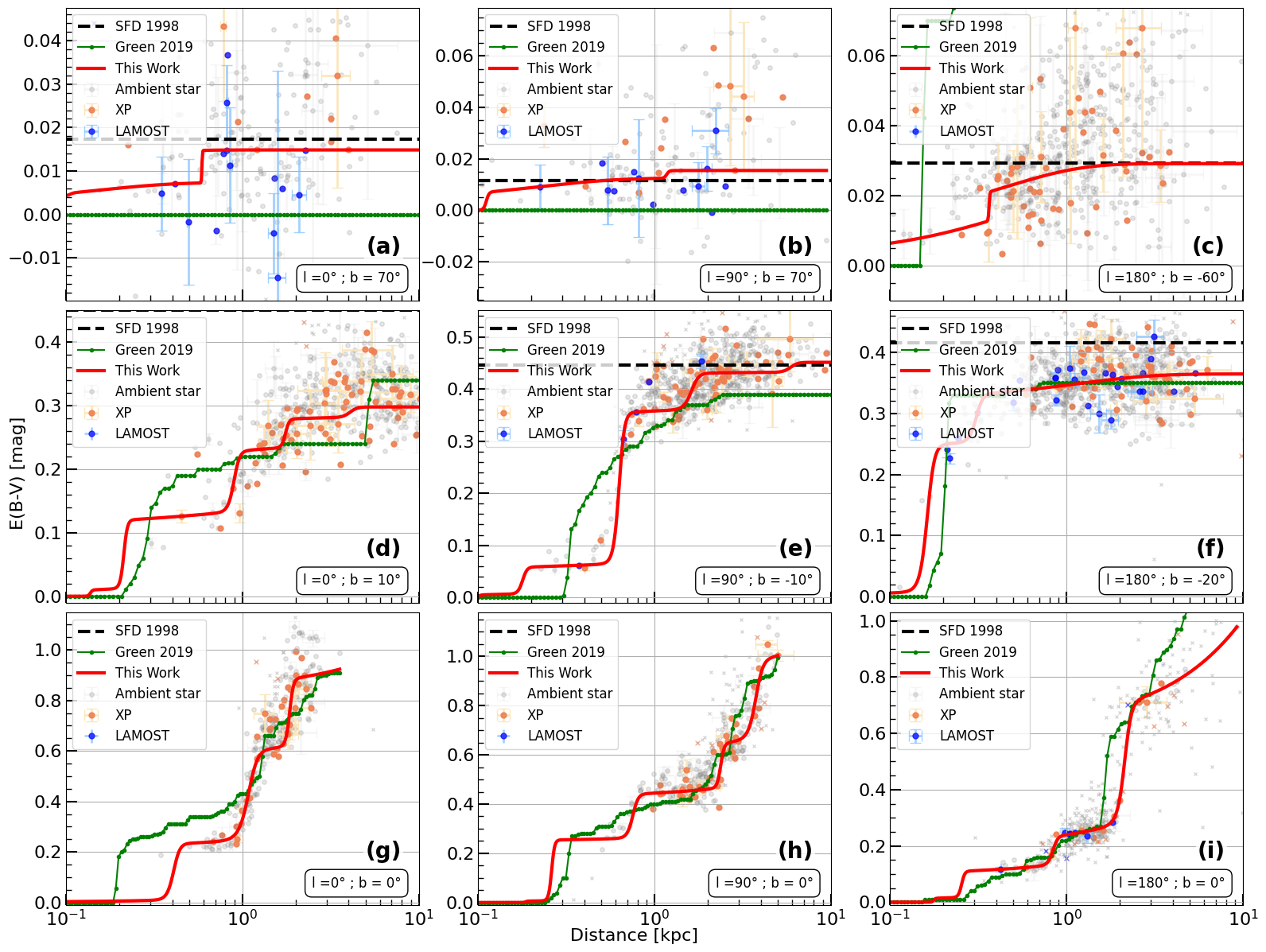}
    \caption{Examples of the distance-extinction relationship in different line-of-sight directions. Blue points represent \ebv$_{\rm LAMOST}$, orange points represent \ebv$_{\rm XP}$, and gray points represent data that are not within the resolution but included with a certain weight.  To avoid overlapping error bars, only 50\% of the blue points, 25\% of the orange points, and 10\% of the gray points have their error bars plotted.  "×"-shaped points in different colors represent data removed after 3-$\sigma$ clipping. The red line is the optimal fitting line of our work. The black dashed line indicates the extinction read from the SFD map in this line-of-sight direction. The green line represents the extinction read from the \cite{Green2019} map.}
    \label{fig:this_work_sfd_green}
\end{figure*}

\begin{figure*}
    \centering
    \includegraphics[width=1\linewidth]{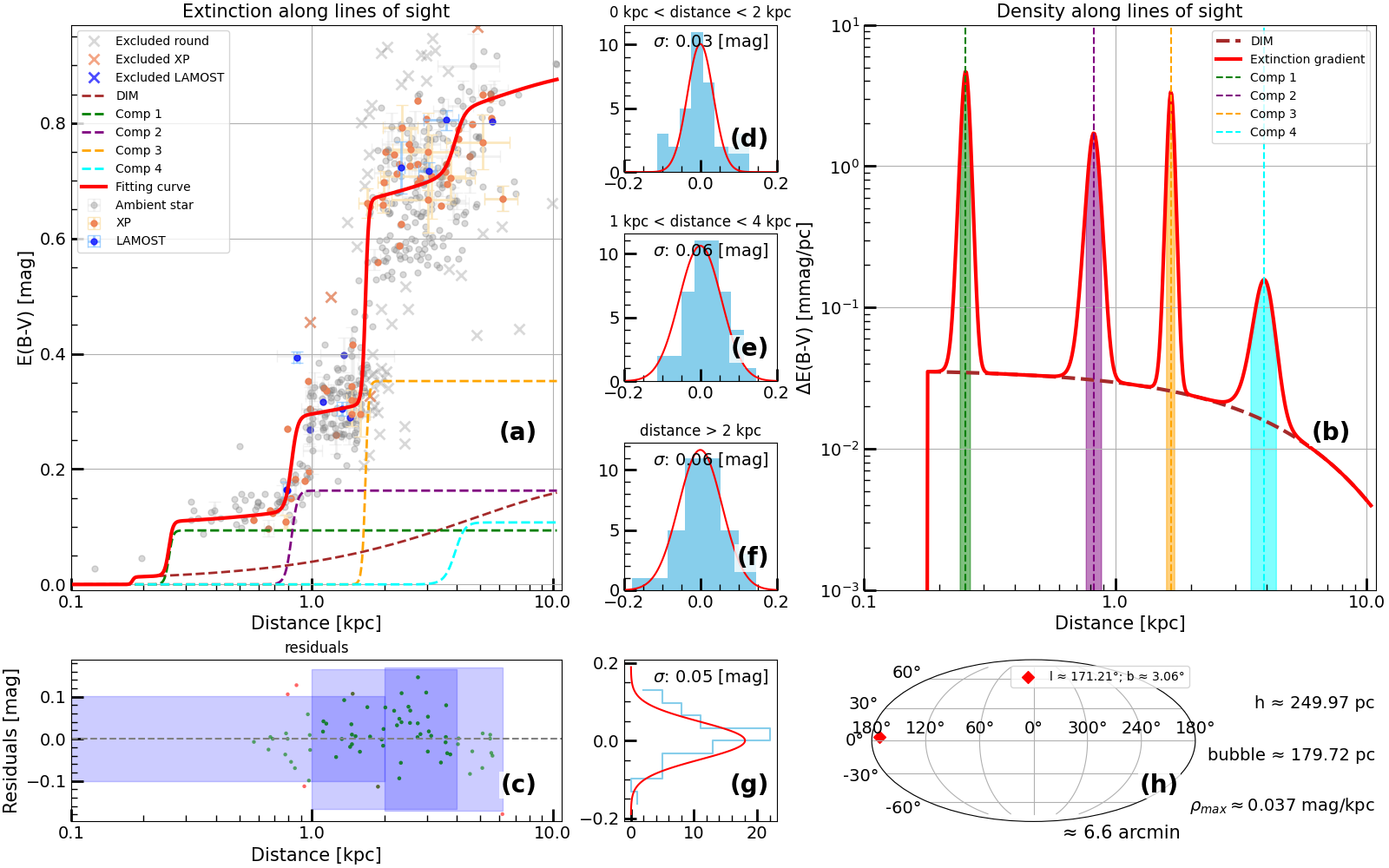}
    \caption{It shows all the information obtained in one line-of-sight. (a) The variation of extinction with distance. Blue points represent \ebv$_{\rm LAMOST}$, orange points represent \ebv$_{\rm XP}$, and gray points represent data that are not within the resolution but included with a certain weight. To avoid overlapping error bars, only 50\% of the blue points, 25\% of the orange points, and 10\% of the gray points have their error bars plotted. "×"-shaped points in different colors represent data removed after 3-$\sigma$ clipping. The red line is the best-fit function for this line-of-sight, and the brown dashed line represents the contribution from diffuse dust. The other dashed lines represent the contributions from different molecular clouds. (b) The variation of dust density with distance. The red line is the derivative of the best-fit line from panel (a), the brown dashed line represents the contribution from diffuse dust, and the shaded regions represent different molecular cloud components. The width of the shading indicates the size of the defined molecular clouds $\Lambda_i^{\mathrm{MC}}$, and the dashed lines mark the distances to the molecular clouds. (c) The residual distribution of the fit shown in panel (a). Only the residuals of the colored points are shown, with blue shading indicating the 3-$\sigma$ range for different distance bins.  The gray dashed line represents a residual of zero. (d), (e), (f) Histograms of the residuals for the distance bins 0-2 kpc, 1-4 kpc, and > 2 kpc, respectively. The red line represents a normal distribution fit, with the $\sigma$ value indicated in the panel. (g) The total residual histogram for this line-of-sight, with the red line representing a normal distribution fit and the sigma value indicated. (h) The position of this line-of-sight in Galactic coordinates, shown as a red square. The resolution of 6.6 arcminutes is marked in the lower right corner. Three parameter values, which are not easily discernible, are also indicated in the lower right corner.
    }
    \label{fig:Details_example}
\end{figure*}

In high Galactic latitude regions, most lines-of-sight require no moren than 2 components to adequately model the reddening (see Figure \ref{fig:this_work_sfd_green} a,b,c). For lines-of-sight that traverse multiple molecular clouds, our model is capable of resolving and separating the characteristics of each individual cloud (see Figure \ref{fig:Details_example} b). The specific contribution of each component to the total reddening effect is visually represented by dashed lines of different colors.

By differentiating the distance-reddening function, we can derive the trend of dust density as a function of distance (See Figure \ref{fig:Details_example} (b)). Overall, this trend exhibits an exponential decline, indicating that the density of diffuse dust decreases with increasing distance. The intercept of this trend with the y-axis represents the maximum dust density at the Sun, while the distance at which the density drops to $1/e$ corresponds to the projected height of the diffuse dust in this line-of-sight. The observed peaks in this process correspond to the positions of different molecular cloud components.  These peaks not only indicate the presence of molecular clouds but also provide key information about their properties.  The area under each peak reflects the cumulative extinction caused by the associated molecular cloud, while the position of the peak along the distance axis reveals the distance to the cloud. Additionally, $\Lambda_i^{\mathrm{MC}}$, as defined in subsection \ref{sec:function} and shown in Figure \ref{fig:Details_example} (with different color shading representing various widths), carries information about the size of the cloud. However, it is important to note that the observed broadening is primarily due to distance uncertainties rather than the intrinsic size of the molecular clouds.

In subsection \ref{sec:fitting}, we applied a 3-$\sigma$ criterion to remove outliers. Subsequently, we defined the distance to the farthest star in each line-of-sight as the maximum reliable distance for that direction (see Figure \ref{fig:HEALPix} (d). Accordingly, the applicable distance extends to approximately 3 -- 5 kpc in the $|b| < 5^\circ$ region, and to 10 -- 15 kpc in the $|b| > 5^\circ$ region.

To assess the fitting quality for each line-of-sight, we computed the residuals' $\sigma$ of the shaded points (see Figure \ref{fig:Details_example}). Additionally, recognizing that measurement errors tend to increase with distance, we not only calculated the overall residual $\sigma$ for each line-of-sight, but also evaluated the $\sigma$ values for different distance ranges separately.

\begin{figure*}
    \centering
    \includegraphics[width=0.49\linewidth]{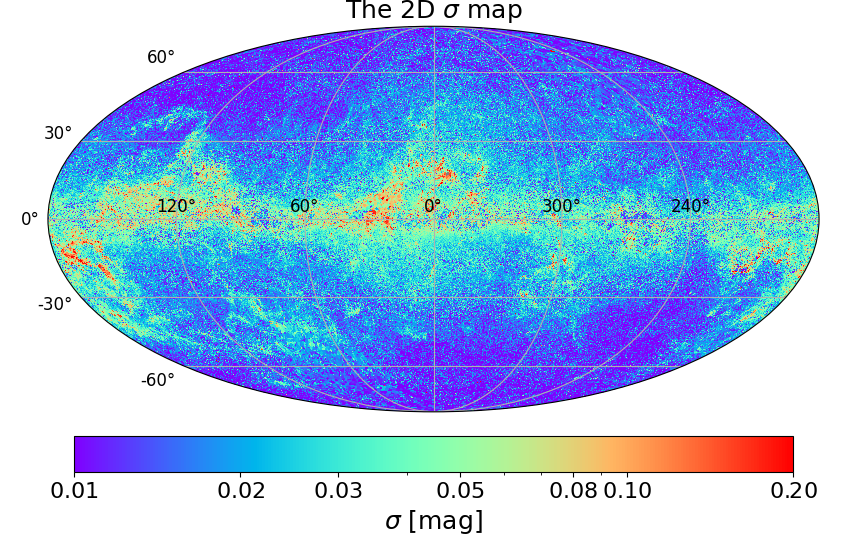}
    \hfill
    \includegraphics[width=0.49\linewidth]{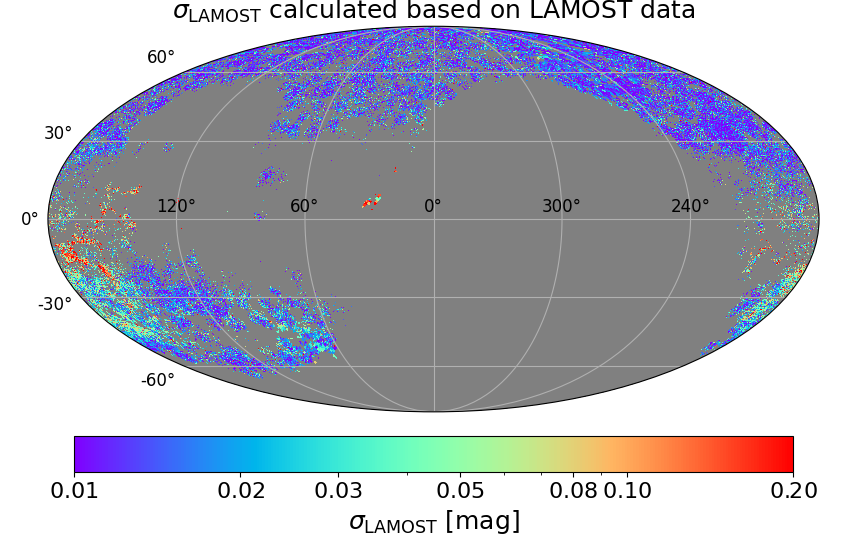}
    \vspace{1em}
    \includegraphics[width=0.95\linewidth]{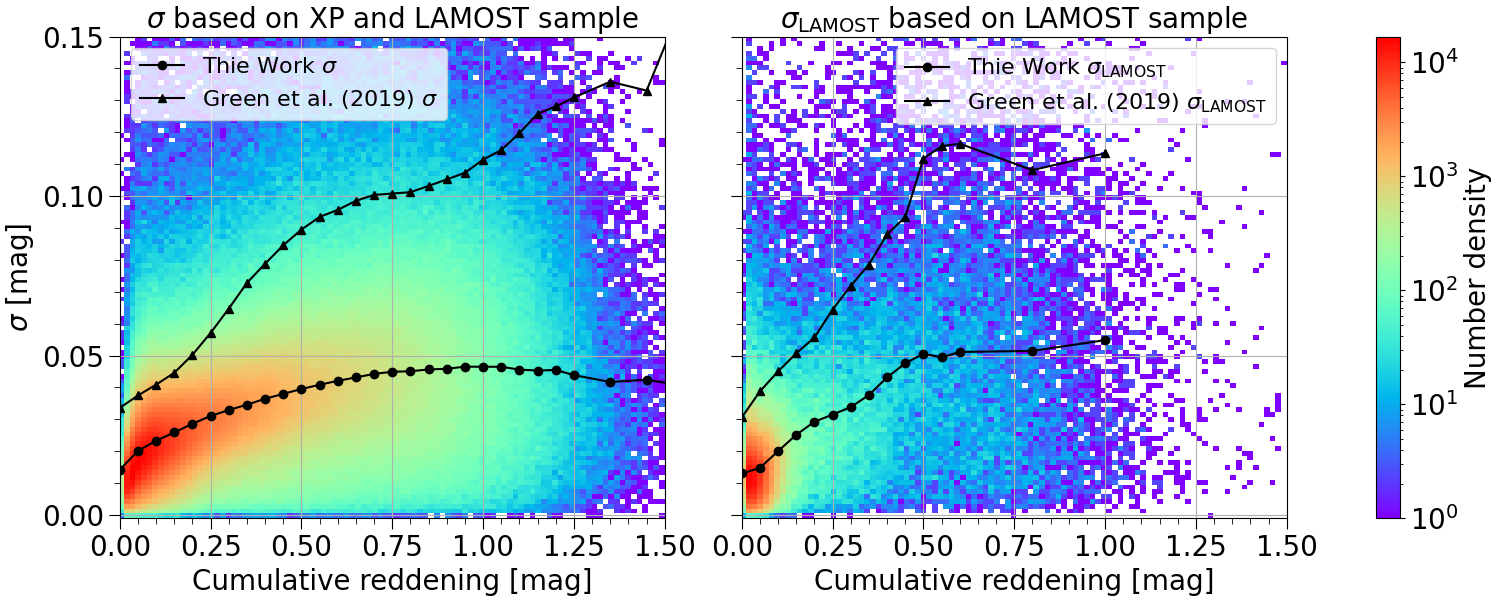}
    \caption{Sky distribution and variation with cumulative extinction of the fitted residual standard deviation ($\sigma$) along different lines of sight. The upper panels show sky maps of $\sigma$ calculated using both LAMOST and XP sources, as well as $\sigma_{\mathrm{LAMOST}}$ calculated using only LAMOST sources. The lower panels present the variation of these $\sigma$ values with cumulative extinction. For comparison, the lower panels also include $\sigma$ values directly calculated from the extinction curves provided by \cite{Green2019}.}
    \label{fig:hp_sigma}
\end{figure*}

In the top panel Figure \ref{fig:hp_sigma}, we present the all-sky distribution of the fitting residual $\sigma$ for different line-of-sight. In the Galactic disk region, typical $\sigma$ derived from both XP and LAMOST data are $\sim$ 0.05 mag. Outside of the disk, the typical $\sigma$ decreases to $\sim$ 0.01 - 0.02 mag. 

In the lower panel of Figure \ref{fig:hp_sigma}, we present the cumulative extinction (defined as the extinction at the point corresponding to the maximum reliable distance) along the line-of-sight and the residuals $\sigma$, with the $\sigma$ values derived from the distance-extinction curves of \cite{Green2019} overplotted for reference. We observe a dependency between $\sigma$ and the cumulative extinction. Moreover, we find that the median of the $\sigma$ computed from the data gradually increases from approximately 0.01 mag to around 0.05 mag before stabilizing. Therefore, we roughly estimate that the precision of our 3D map is between approximately 0.01 mag and 0.05 mag, attaining around 0.01 mag across most high Galactic latitude regions ($|b| > 20^\circ$), while it degrades to approximately 0.01 -- 0.05 mag at low Galactic latitudes ($|b| < 20^\circ$).

\subsection{Comparisons with Previous Work} \label{sec:Comparisons}

\subsubsection{Schlegel et al. (1998)} \label{sec:SFD}

\begin{figure}
    \centering
    \includegraphics[width=1\linewidth]{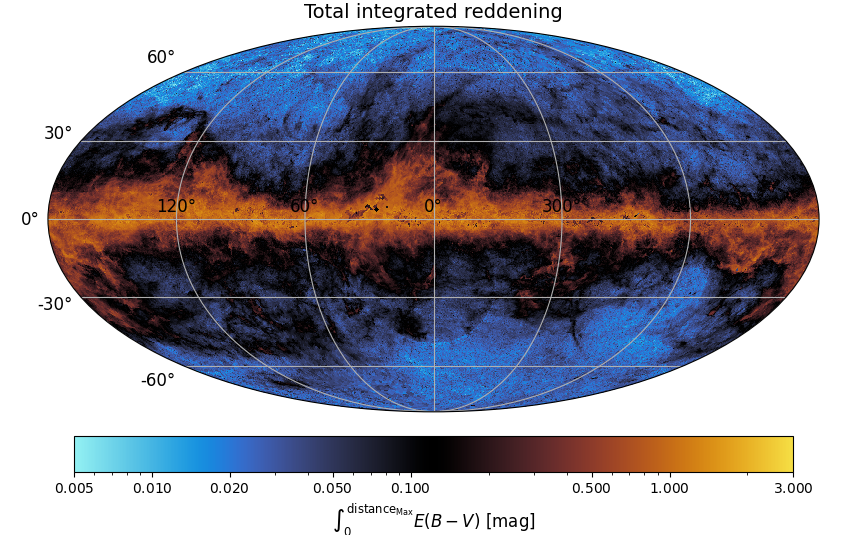}
    \includegraphics[width=1\linewidth]{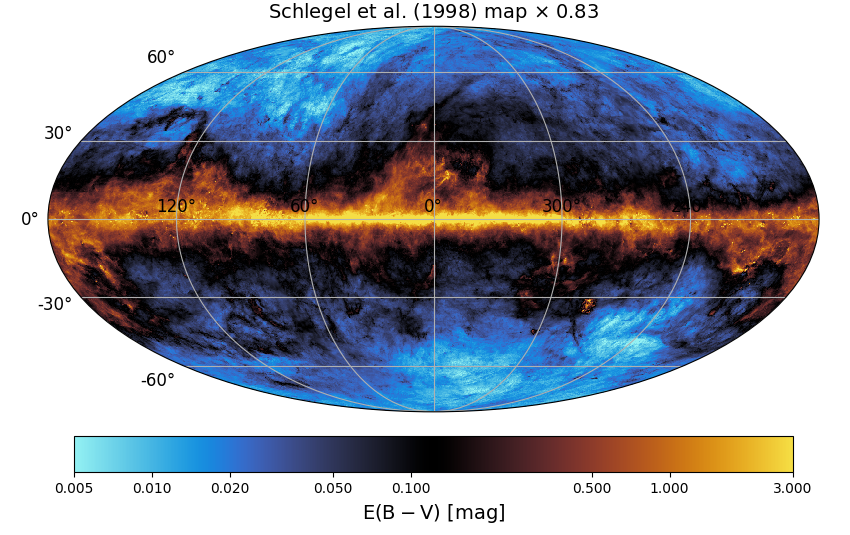}
    \includegraphics[width=1\linewidth]{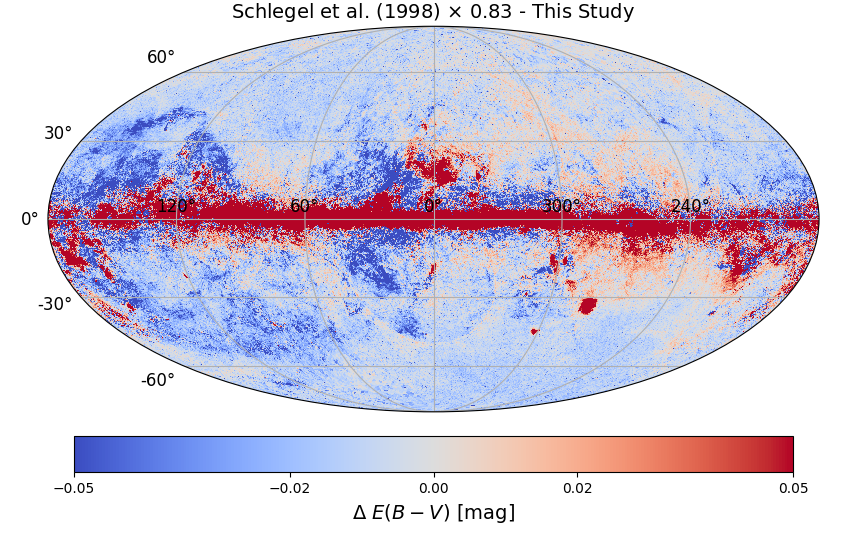}
    \caption{Comparison to the 2D extinction map of \cite{SFD1998}. Top panel: Our 3D extinction map at the maximum reliable distance, with E(B$-$V) shown at a resolution of 3.3 arcmin, corresponding to $\rm N_{side}$ = 1024. Middle panel: The extinction map from \cite{SFD1998} is scaled by a factor of 0.834, also shown at a resolution of 3.4 arcmin ($\rm N_{side}$ = 1024). Bottom panel: The difference between the extinction map in the middle panel and the map in the top panel.}
    \label{fig:sfd_vs_our}
\end{figure}

\begin{figure}
    \centering
    \includegraphics[width=1\linewidth]{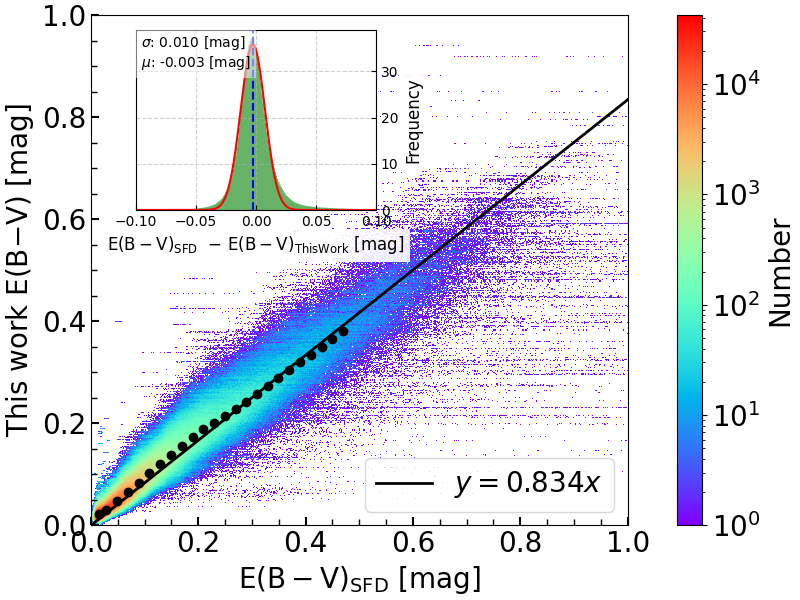}
    \caption{A comparison of the E(B$-$V) derived from the extinction integral at the maximum reliable distance in this work and those from the SFD is shown. The black points represent the median values within each bin, and the black line is the linear fit to these points. \st{Note that regions with |b| > 20$^\circ$ and MCs have been excluded.} LMC, SMC, and |b| < 20$^\circ$ regions were excluded.} The upper-left subplot displays a histogram of the differences between the extinction from this work at the maximum reliable distance and those from the SFD, along with a normal distribution function fit.
    \label{fig:sfd_our}
\end{figure}

The SFD map, derived from all-sky far-infrared data, provides the integrated extinction along the line-of-sight. To compare our measurement results with the SFD map, we selected the extinction at the maximum reliable distance for each line-of-sight and plotted them in Figure \ref{fig:sfd_vs_our} (a). Additionally, to account for systematic differences between our work and the SFD map, we examine the extinction ratio at Galactic latitudes with absolute values greater than 20$^\circ$ (see Figure \ref{fig:sfd_our}), revealing a scaling factor of 0.834. This indicates that the SFD results are approximately 16\% higher than our findings,  consistent with \cite{2011ApJ...737..103S} and \cite{2013MNRAS.430.2188Y}.

In Figure \ref{fig:sfd_our}, we apply this scaling factor to the SFD map and compare the scaled map with our integrated extinction map. The difference between the Galactic disk, LMC and SMC is both evident and easily understood, as substantial amounts of dust still exist beyond the maximum reliable distance of our observations. The differences in the high Galactic latitude region are consistent with the conclusions drawn in Figure 10 of \cite{2022ApJS..260...17S}, but our results reveal more detailed structures due to the wider sky coverage and higher angular resolution of our data. 

\begin{figure}
    \centering
    \includegraphics[width=1\linewidth]{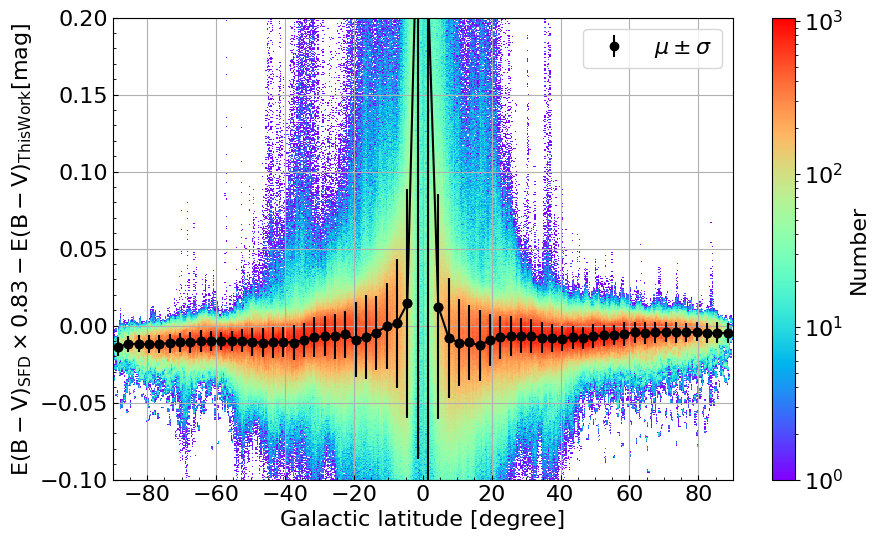}
    \caption{The difference with the \cite{SFD1998}, after scaling by a factor of 0.834, is shown in the 2D histogram of the Galactic latitude. The black points represent the mean values calculated after a 3-$\sigma$ clipping within each Galactic latitude bin, and the error bars indicate the standard deviation.}
    \label{fig:sfd_GL}
\end{figure}

We compare the results from our study with those from the SFD map as a function of Galactic latitude, as shown in Figure \ref{fig:sfd_GL}.  Although this comparison has its limitations, it is clear that there is good agreement between our results and the SFD map at high Galactic latitudes.

\subsubsection{Green et al. (2019)} \label{sec:Green}

\begin{figure*}
    \centering
    \includegraphics[width=1\linewidth]{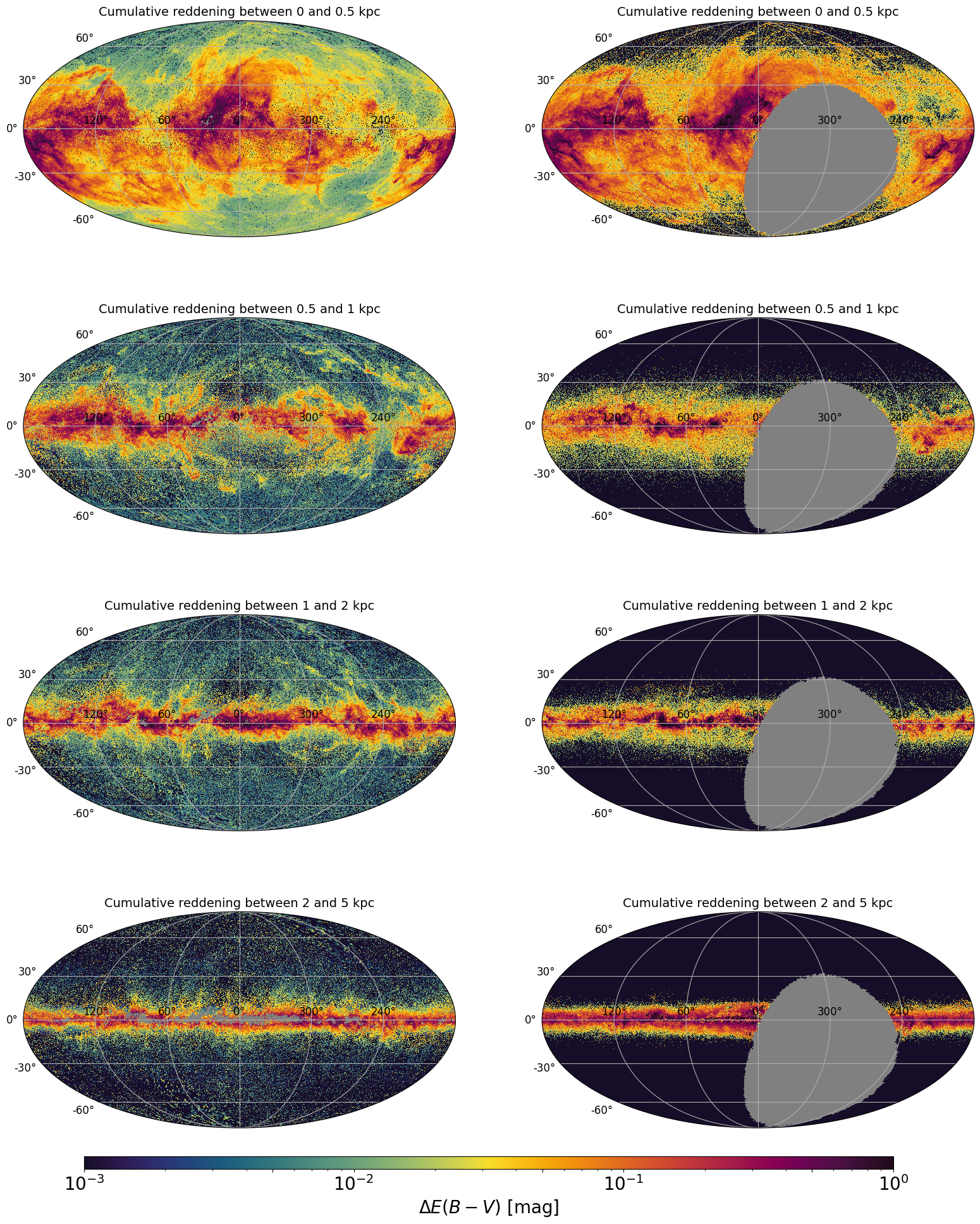}
    \caption{Cumulative reddening in the Galactic Mollweide projection at different distances: 0 - 0.5 kpc, 0.5 - 1 kpc, 1 - 2 kpc, and 2 - 5 kpc. The left side of the figure shows the results from our work, while the right side displays the results from \cite{Green2019}. All subplots are plotted with a resolution of 3.3 arcminutes, corresponding to $\rm N_{side}$ = 1024.}
    \label{fig:green_our_4fig}
\end{figure*}

In Figure \ref{fig:green_our_4fig}, we compare the cumulative extinction over different distance intervals from our work with the 3D extinction map provided by \cite{Green2019}. The distance intervals are 0 -- 0.5 kpc, 0.5 -- 1 kpc, 1 -- 2 kpc, and 2 -- 5 kpc. We find that, on a large scale, the two datasets show good agreement. However, our work reveals more intricate structural details. Furthermore, our work uncovers previously unnoticed details: within the 0.5 -- 1 kpc range, prominent molecular clouds are present in the high Galactic latitude regions; and within the 0.5 -- 2 kpc range, structures extending from the Galactic disk are clearly visible. We will investigate these structures further in our subsequent work.

We observe that in most regions where |b| > 60$^\circ$ , \cite{Green2019} report zero extinction (see Figure 15a and 15b), whereas our data suggest that some extinction should be present.  Additionally, \cite{Green2019} show similar results in certain low Galactic latitude line-of-sight directions (see the black regions in their Figure 5). At the same time, in the majority of high-extinction regions, our distance-extinction curve aligns relatively well with that of \cite{Green2019}, but there are still significant and non-negligible discrepancies in some details (see Figure 15d, e, g, and f).

\subsection{Dust distribution} \label{sec:Dust}

\begin{figure*}
    \centering
    \includegraphics[width=1\linewidth]{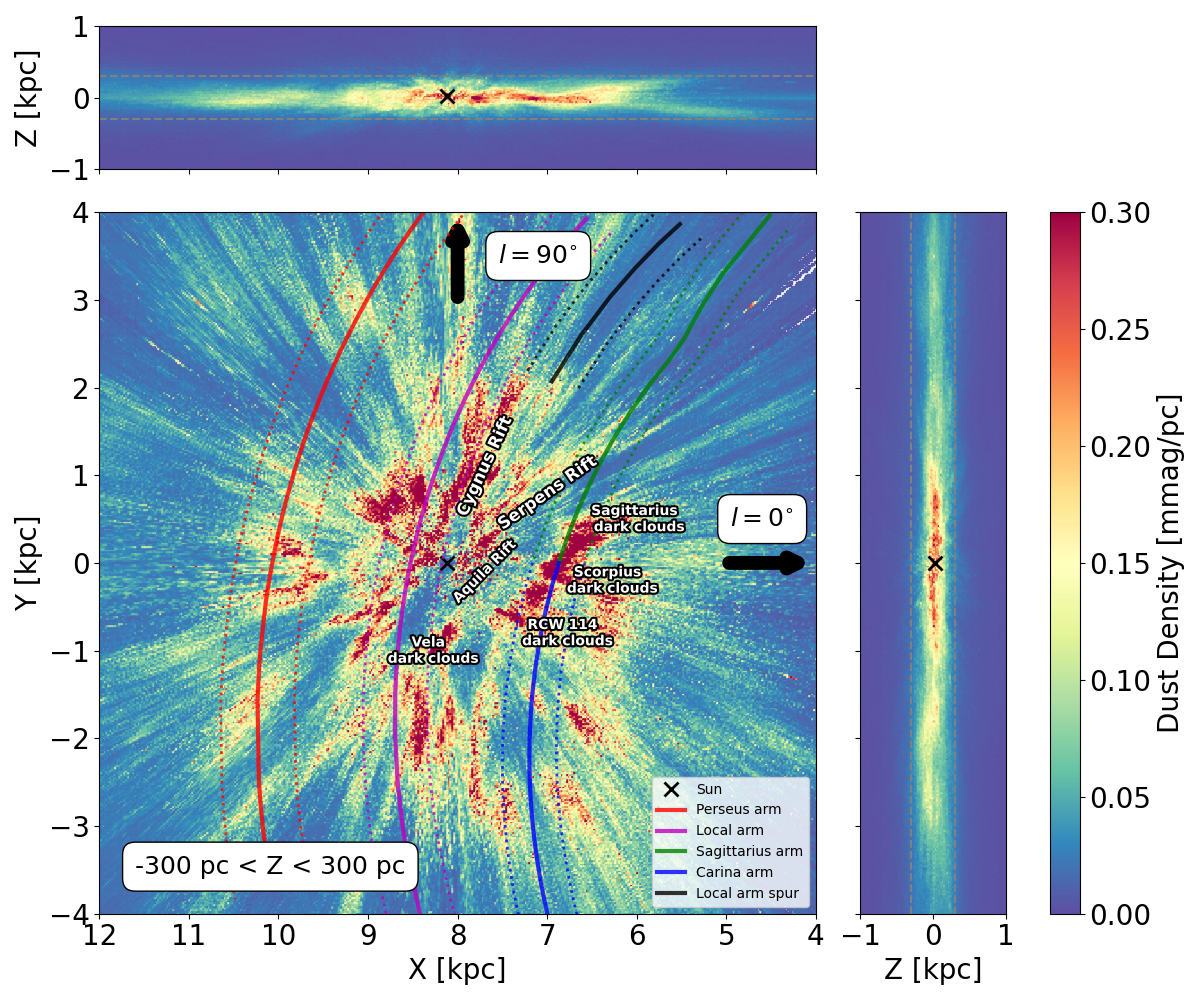}
    \caption{Dust density in the XY plane containing the Sun, with the Sun located at a distance of 8.12 kpc from the Galactic center (marked by the black "×" symbol). The directions of Galactic longitude $l = 0^\circ$ and $l = 90^\circ$ are also indicated. The solid black and dashed curved lines represent the center and $\pm 1 \sigma$ widths of the spiral arm models, from left to right: Perseus arm, Local arm, Local arm spur, and Sagittarius-Carina arm, as taken from \cite{2019ApJ...885..131R} cantilever position obtained using proper motions of molecular masers. Top: Dust density in the XZ plane. Right: Dust density in the YZ plane.}
    \label{fig:Disk_dust}
\end{figure*}

In Figure \ref{fig:Disk_dust}, we show the Galactic XY, YZ, and XZ planes, with the Sun located at X = 8.12 kpc, Y = 0 kpc and Z = 0.02 kpc. From the YZ and XZ planes, we observe that the disk dust in the Galactic disk is concentrated within |Z| < 300 pc. Therefore, we plot the differential reddening, $\Delta$ \ebv, in the Galactic disk within the region |Z| < 300 pc, expressed in units of mag/kpc. The overall morphology of the dust structure matches well with those of Figure 12 in \cite{2019MNRAS.483.4277C}, Figure 12 in \cite{Green2019}, Figure 10 in \cite{2022AA...661A.147L}, and Figure 11 \cite{2022AA...664A.174V}.

In Figure \ref{fig:Disk_dust}, several discrete molecular clouds are clearly associated with the Perseus Arm. Dust-dense regions within the Local Arm, such as the Aquila Rift, Cygnus Rift, and Serpens Rift, are particularly prominent and distinctly visible. Additionally, discrete molecular clouds including the Sagittarius Dark Cloud, Scorpius Dark Cloud, and RCW 114 Dark Cloud are evidently linked to the Sagittarius-Carina Arm. However, a detailed investigation of these structures lies beyond the scope of this paper.

\section{Easy to use}  \label{sec-specification}

To facilitate easy access to the aforementioned data, we have developed a dedicated website (\url{https://nadc.china-vo.org/data/dustmaps/}) and a Python package (available on GitHub) designed to allow users to quickly retrieve all relevant information. Our approach utilizes a parametric design, ensuring that the Python package is compact ($\sim$ 350 MB) while maintaining fast data access speeds. Users need only input the three-dimensional coordinates to obtain a set of parameters, including:

\begin{enumerate}
    \item Extinction (\ebv, in mag),
    \item Dust density (i.e., the extinction gradient, in mag/kpc);
    \item Uncertainty (i.e., the $\sigma$ values discussed in subsection \ref{sec:map});
    \item The maximum reliable distance along the given line-of-sight;
    \item The distance to the local bubble along the line-of-sight;
    \item The height of diffuse dust along the line-of-sight;
    \item The number of molecular clouds along the line-of-sight;
    \item Whether the given position lies within a molecular cloud, and if so, detailed parameters of the cloud.
\end{enumerate}

The tool can be quickly installed via \texttt{pip install dustmaps3d}. The usage documentation can be viewed at \url{https://github.com/Grapeknight/dustmaps3d/}.

Once the E(B$-$V) values are obtained from our 3D dust map, users may wish to apply extinction corrections or reddening adjustments in specific photometric bands. This requires the use of reddening coefficients to convert E(B$-$V) into extinction in specific bands or reddening for color indices. For example, \citet{2023ApJS..264...14Z} provide reddening coefficients that account for variations with stellar effective temperature and extinction. These coefficients are available in Tables 2--4 of their paper and are implemented in the Python package \texttt{extinction\_coefficient}\footnote{The reddening coefficients from \citet{2023ApJS..264...14Z} are calibrated against the SFD map. Therefore, to use them with our E(B$-$V), a scaling factor of $1/0.834$ should be applied to correct for the systematic offset between our map and the SFD map (see Figure~\ref{fig:sfd_our}).} (\url{https://github.com/vnohhf/extinction_coefficient}
). However, it is important to note that these coefficients are calibrated only for the extinction range $E(B-V) \lesssim 0.5$ mag, as the training sample used by \citet{2023ApJS..264...14Z} did not include stars with higher extinction. Therefore, their applicability is limited in regions of high dust extinction.

For photometric bands or color indices not included in \citet{2023ApJS..264...14Z}, or when the extinction exceeds $E(B-V) > 0.5$ mag, we recommend using the more general \texttt{XP\_Extinction\_Toolkit}\footnote{When using the \texttt{XP\_Extinction\_Toolkit} (\url{https://github.com/vnohhf/XP_Extinction_Toolkit}
) provided by \citet{2024ApJ...971..127Z}, the E(B$-$V) from this work can be directly used as E(440--550) inputs without any scaling.} developed by \citet{2024ApJ...971..127Z}. This toolkit offers greater flexibility and broader applicability, providing functions such as \texttt{star\_ext} and \texttt{star\_reddening}, which compute reddening coefficients based on the median Galactic extinction curve derived in their study. Given stellar parameters (T$_\mathrm{eff}$, $\log g$), filter transmission curves, and the input extinction E(440--550), the toolkit can estimate reddening coefficients for arbitrary bands, making it suitable for both uncommon filters and high-extinction environments.

\section{Summary} \label{sec-Summary}

We utilized stellar atmospheric parameters from LAMOST DR11 and successfully obtained \ebv$_{\rm LAMOST}$ for approximately 4.6 million unique sources by combining the standard-pair algorithm with the synthetic B and V-band magnitudes from $\gaia$. The typical precision of these measurements reaches $\sim$ 0.01 mag. 

In addition, from \href{https://doi.org/10.5281/zenodo.7692680}{Z23}, we selected \ebv$_{\rm XP}$ for approximately 150 million stars from a catalog of 220 million parameterized stars, using \ebv${\rm LAMOST}$ from LAMOST as the reference. For \ebv$_{\rm XP}$, the typical precision for common sources between XP and LAMOST is approximately 0.015 mag, with an overall median extinction precision of around 0.03 mag. We then merged the extinction values from LAMOST and XP and incorporated the revised distances provided by \cite{zhang2023}, thereby forming our final dataset.

Subsequently, we adaptive divided the whole sky into five lines-of-sight with different resolutions, selected the stars in the line-of-sight as the main probe, and selected the ambient stars to participate in the modeling with low weight. By establishing a distance-extinction relationship model consisting of a "local bubble + diffuse dust + n possible molecular clouds," we fitted the extinction data of these stars to derive the distance-extinction relations for each line-of-sight direction, ultimately constructing a 3D extinction map.

This 3D extinction map covers the entire sky, providing reliable coverage up to 3 -- 5 kpc in the $|b| < 5^\circ$ region and up to 10 -- 15 kpc at $|b| > 5^\circ$. The angular resolution ranges from 3.3 to 54 arcminutes, with about half of the sky having a resolution better than 6.9 arcminutes. The extinction precision reaches approximately 0.01 -- 0.05 mag, attaining around 0.01 mag across most high Galactic latitude regions ($|b| > 20^\circ$), while it degrades to approximately 0.01 -- 0.05 mag at low Galactic latitudes ($|b| < 20^\circ$). The map is publicly available on the website \url{https://nadc.china-vo.org/data/dustmaps/} and can also be accessed via the Python package \texttt{dustmaps3d}. For applications requiring extinction corrections in different photometric bands or reddening estimates for various color indices, the reddening coefficients provided by \citet{2023ApJS..264...14Z} and \citet{2024ApJ...971..127Z} can be used for the necessary transformations. Additionally, the LAMOST extinction data used in this study is also published on \url{https://nadc.china-vo.org/res/r101617/}.

Our integrated reddening estimates reveal that the SFD maps overestimate reddening by approximately 16\%, consistent with the findings of \citet{2010ApJ...725.1175S}, \citet{2011ApJ...737..103S}, and \citet{2013MNRAS.430.2188Y}. In regions of the MW with high Galactic latitudes, our integrated reddening estimates show strong agreement with SFD. However, the discrepancies between our estimates and those of SFD exhibit complex spatial dependence patterns, aligning with the results reported by \cite{2022ApJS..260...17S}. In comparison with \cite{Green2019}, we identified previously overlooked details, specifically that regions of high reddening density extend outward along the Galactic disk. The spiral arm structure of the MW remains not fully elucidated, with multiple models proposed to explain its configuration. \cite{2019ApJ...885..131R} utilized the proper motions of molecular masers to trace the spiral arms, and we find that our regions of high reddening density are associated with the Perseus Arm and the Sagittarius-Carina Arm as identified by \cite{2019ApJ...885..131R}.

The 3D reddening map we developed demonstrates extensive potential in two key areas: precise extinction correction and the global morphology of the Milk Way and its ISM. First, in the domain of precise extinction correction, this 3D reddening map provides indispensable support for various astronomical investigations. It is crucial for correcting extinction of objects embedded in the Galactic disk, generating high-precision photometric standards stars, measuring stellar parameters, determining the period-luminosity relationships of Cepheid variables, and establishing the absolute intrinsic luminosities of horizontal branch stars. In the study of the global morphology of the Milk Way and its ISM, the parameterized approach that we employed offers significant advantages in analyzing bubbles, DIM, and molecular clouds. This method not only facilitates the identification and distance measurement of molecular clouds (\citeauthor{Wang2025b} submitted) but also serves as a powerful tool for validating neutral hydrogen associations, confirming supernova remnants, analyzing the structures of local bubbles, and investigating the architecture of the diffuse interstellar medium. In future work, we will continue to expand our work using the 3D reddening map and anticipate discovering numerous applications that have not yet been foreseen. With the continuous enrichment and improved precision of extinction and distance data in the future, our 3D dust reddening map could be updated in a convenient and efficient way.

\begin{acknowledgments}
The authors thank the anonymous referee for his/her suggestions that improved the quality and clarity of our presentation.
The authors acknowledge helpful discussions with Xiangyu Zhang.
This work is supported by the National Natural Science Foundation of China through the projects NSFC 12222301, 12173007, and 12173034, as well as
the National Key Basic R\&D Program of China via 2024YFA1611901 and
2024YFA1611601.  
This work is also supported by the China Manned Space Program
with grant no. CMS-CSST-2025-A13.

This work has made use of data products from the LAMOST and $\gaia$. 
Guoshoujing Telescope (the Large Sky Area Multi-Object Fiber Spectroscopic Telescope; LAMOST) is a National Major Scientific Project built by the Chinese Academy of Sciences. 
Funding for the project has been provided by the National Development and Reform Commission. 
LAMOST is operated and managed by the National Astronomical Observatories, Chinese Academy of Sciences. 

\end{acknowledgments}

\clearpage
\bibliography{ref.bib}

\end{CJK*}
\end{document}